\documentclass[12pt]{article}
\usepackage{a4wide}
\usepackage{fullpage}
\usepackage{slashbox}
\usepackage{tabularx} 

\usepackage{times}
\usepackage{epsf}
\usepackage{amsmath,amsfonts,amssymb}
\usepackage{graphicx}
\usepackage{array}

\newcounter{definition}

\def\thefigure{\arabic{section}.\arabic{figure}}
\def\theequation{\thesection.\arabic{equation}}
\def\appendix{
  \setcounter{section}{0}
  \setcounter{subsection}{0}
  \par
  \def\thesection{Appendix \Alph{section}}
  \def\theequation{\Alph{section}.\arabic{equation}}
  \def\thefigure{\Alph{section}.\arabic{figure}}
}

\def\d{\delta}

\def\l{\lambda}

\def\e{\epsilon}

\def\be{\begin{equation}}
\def\ee{\end{equation}}
\def\beq{\begin{align}}
\def\eeq{\end{align}}
\def\ov{\overline}

\def\be{\begin{equation}}
\def\ee{\end{equation}}
\def\beq{\begin{eqnarray}}
\def\eeq{\end{eqnarray}}
\def\ov{\overline}

\def\({\left(}
\def\){\right)}
\def\[{\left[}
\def\]{\right]}
\def\bra{\langle}
\def\ket{\rangle}

\def\debut{\begin{align}}
\def\fin{\end{align}}

\def\l{\lambda}
\def\e{\epsilon}
\title{
First principle approach to correlation functions of spin-1/2 
Heisenberg chain : fourth-neighbor correlators
} 
\author{
  H.~E. {\sc Boos}${}^1$\thanks{boos@physik.uni-wuppertal.de}, \ \
  M. {\sc Shiroishi}${}^2$\thanks{siroisi@issp.u-tokyo.ac.jp}, \ \ 
  M. {\sc Takahashi}${}^2$\thanks{mtaka@issp.u-tokyo.ac.jp}
\\
 \\ \it
   ${}^1$ Physics Department, University of Wuppertal, D-42097,
Wuppertal, Germany\footnote{
on leave of absence from the Institute for High Energy Physics, 
Protvino, 142281, Russia}\\\it
  ${}^2$Institute for Solid State Physics, University of Tokyo,\\\it 
  Kashiwanoha 5-1-5, Kashiwa, Chiba 277-8581, Japan\\\it
}

\begin{document}
\maketitle
\setlength{\baselineskip}{1.8em}
\begin{abstract}
We show how correlation functions of the 
spin-1/2 Heisenberg chain without magnetic field in the 
anti-ferromagnetic ground state can be explicitly calculated 
using information contained in the quantum Knizhnik-Zamolodchikov 
equation [qKZ]. 
We find several fundamental relations 
which the inhomogeneous correlations should fulfill. 
On the other hand, it turns out that
these relations can fix the form of the correlations uniquely. 
Actually, applying this idea, 
we have obtained all the correlation functions on five sites. 
Particularly
by taking the homogeneous limit, we have got the analytic form of the 
fourth-neighbor pair correlator ${\bra S_j^z S_{j+4}^z \ket}$.
\end{abstract}

\newpage

\noindent

\section{Introduction}
\setcounter{equation}{0}
The anti-ferromagnetic spin-1/2 Heisenberg ${XXX}$ chain \cite{Heis}
\begin{equation}
H= \sum_{j=1}^{L}\(S_j^xS_{j+1}^x+S_j^yS_{j+1}^y+S_j^zS_{j+1}^z\),
\label{Hamiltonian}
\end{equation} 
is one of the most fundamental models in the study of quantum 
magnetism in low dimensions (see, for instance, the book
\cite{takbook}).

The Hamiltonian (\ref{Hamiltonian}) was diagonalized by Bethe in 1931 
\cite{B} and the ground state in the thermodynamic limit 
was investigated by Hulth\'{e}n in 1938 \cite{H}. The physical content
was clarified by Faddeev and Takhtajan \cite{FT} using the
algebraic Bethe ansatz formulated by Faddeev, Sklyanin and
Takhtajan (see the review \cite{ABA}). Exact integrability of
the Heisenberg model is intimately connected with the Yang-Baxter
equation \cite{bax}.

The Heisenberg model also appears as an effective Hamiltonian of more 
complicated systems such as the one-dimensional Hubbard model at half-filling with 
a large coupling constant ${U}$. 

Since discovering of the model a lot of physical quantities have been 
calculated exactly which afford significant and reliable data for 
the experiments as well as numerical simulations. 
However, most of them are the bulk quantities 
and we have obtained little results on the correlation functions. 
Especially significant are 
the (static) spin-spin pair correlators $\bra S_j^z S_{j+k}^z \ket$ 
(or equivalently 
$\bra S_j^+ S_{j+k}^-\ket/2$;  $S_j^{\pm}=S_j^x\pm i S_j^y$), 
for which only the first and the second neighbor ($k=1,2$) have been 
known for a long time: 
\begin{align}
\bra S_j^{z} S_{j+1}^z\ket &=\frac{1}{12}-\frac{1}{3}\ln2
\simeq-0.14771572685, \label{1st} \\
\bra S_j^{z} S_{j+2}^z\ket &=\frac{1}{12}-\frac{4}{3}\ln2+
\frac{3}{4}\zeta(3)\simeq0.06067976996.
\label{2nd}
\end{align}
%
Here $\zeta(s)$ is the Riemann zeta function and 
$\bra \cdots \ket$ denotes the ground state expectation 
value. Note that (\ref{1st}) was derived directly 
from the ground state energy in the thermodynamic limit by Hulth\'{e}n, 
while (\ref{2nd}) was obtained by Takahashi in 1977 via the strong 
coupling expansion for the ground state energy of the half-filled 
Hubbard model \cite{tak} 
(see also another derivation by Dittrich and Inozemtsev 
\cite{DI}). Unfortunately we cannot 
generalize these methods so as to calculate further correlations 
$\bra S_j^{z} S_{j+k}^z \ket_{k \ge 3}$.

On the other hand, in 1990s, correlation functions were studied systematically
for the Heisenberg ${XXZ}$ model. It was mainly developed by 
Kyoto group (Jimbo, Miwa, Miki, Nakayashiki, \cite{JMN}, see also the
book \cite{JMbook} ). 
Utilizing the representation theory of the quantum affine algebra 
$U_q(\widehat{sl_2})$, 
they first derived multiple integral representation of arbitrary 
correlators for for the massive 
${XXZ}$ anti-ferromagnet. In 1996 Jimbo and Miwa extended
the above results to the ${XXX}$ chain and the massless ${XXZ}$ 
chain \cite{JM}. The manifest integral formula for special case of 
correlation functions, Emptiness Formation Probability (EFP)
of the ${XXX}$ model was earlier obtained by Korepin, Izergin, Essler and
Uglov in \cite{KIEU}.  
Strictly speaking the formulas in the massless region were a 
conjecture since they were obtained 
as solutions to the quantum Knizhnik-Zamolodchikov equations 
(qKZ equations) with a certain assumption. 
However, the same integral formulas have been reproduced by 
Kitanine, Maillet and Terras, \cite{Maillet1, Maillet2} in 
the framework of the Quantum Inverse Scattering Method. 
So we have now at hand the multiple integral 
representations for any arbitrary correlation functions for 
the Heisenberg ${XXZ}$ chain. 

In 2001, Boos and Korepin invented a method to evaluate these multiple 
integrals 
for the Heisenberg ${XXX}$ chain \cite{bk1,bk2}. They have shown that 
the integrand can be reduced to a certain canonical 
form so that the integrations can be performed. 
They applied the method to the case of the Emptiness Formation Probability.
There has been also suggested conjecture that the correlation 
functions for the ${XXX}$ model may be expressed in terms of the
Riemann's zeta function at odd arguments with rational coefficients.
For instance, the results for ${P(4)}$ and ${P(5)}$ look as follows: 
\begin{align}
P(4) & \equiv \bra \left(1/2+S_j^{z}\right) 
\left(1/2+S_{j+1}^{z}\right) 
\left(1/2+S_{j+2}^{z}\right) \left(1/2+S_{j+3}^{z}\right) \ket 
\nonumber \\
&= \frac{1}{5} - 2 \ln 2  + \frac{173}{60} \zeta(3) - \frac{11}{6} \ln 2 
\cdot \zeta(3) 
- \frac{51}{80} \zeta(3)^2 - \frac{55}{24} \zeta(5) + \frac{85}{24} 
\ln 2 \cdot \zeta(5)  \nonumber  \\
& \\
P(5) & \equiv \bra \left(1/2+S_j^{z}\right) 
\left(1/2+S_{j+1}^{z}\right) 
\left(1/2+S_{j+2}^{z}\right) \left(1/2+S_{j+3}^{z}\right) 
\left(1/2+S_{j+4}^{z}\right) \ket \nonumber \\
&= \frac{1}{6} - \frac{10}{3} \ln 2  + \frac{281}{24} \zeta(3) - 
\frac{45}{2} \ln 2 \cdot \zeta(3) 
- \frac{489}{16} \zeta(3)^2 - \frac{6775}{192} \zeta(5) + 
\frac{1225}{6} \ln 2 \cdot \zeta(5) \nonumber  \\
&  - \frac{425}{64} \zeta(3) \cdot \zeta(5) - \frac{12125}{256} 
\zeta(5)^2 + \frac{6223}{256} \zeta(7) - \frac{11515}{64} \ln 2 
\cdot \zeta(7) 
+  \frac{42777}{512}  \zeta(3) \cdot \zeta(7)  \label{P5}
\end{align} 
The last result was obtained in \cite{BKNS}.
 
More recently the method was applied to calculate other correlation 
functions of the Heisenberg chain (\ref{Hamiltonian}).  
Here let us list the results in \cite{SSNT} where all the 
independent correlation functions on four lattice 
sites are obtained:
\begin{align}
\left\langle S_{j}^{z} S_{j+3}^{z} \right\rangle 
& = \frac{1}{12} - 3 \ln 2 + \frac{37}{6} \zeta(3) - \frac{14}{3} \ln 2 
\cdot \zeta(3) 
- \frac{3}{2} \zeta(3)^2  - \frac{125}{24} \zeta(5) + \frac{25}{3} 
\ln 2 \cdot \zeta(5)  \nonumber \\ 
& = -0.05024862725 \cdots, \nonumber \\ 
\left\langle S_{j}^{x} S_{j+1}^{x} S_{j+2}^{z} S_{j+3}^{z} 
\right\rangle 
& = \frac{1}{240}+\frac{1}{12}\ln 2-\frac{91}{240}\zeta(3) 
+ \frac{1}{6} \ln 2 \cdot \zeta(3) + \frac{3}{80} \zeta(3)^2 
+ \frac{35}{96}\zeta(5)-\frac{5}{24} \ln 2 \cdot \zeta(5) \nonumber \\
\left\langle S_j^{x} S_{j+1}^{z} S_{j+2}^{x} S_{j+3}^{z} \right\rangle 
& = \frac{1}{240}-\frac{1}{6}\ln 2+\frac{77}{120}\zeta(3) 
- \frac{5}{12} \ln 2 \cdot \zeta(3)-\frac{3}{20} \zeta(3)^2 -\frac{65}{96}\zeta(5) 
+ \frac{5}{6} \ln 2 \cdot \zeta(5) \nonumber  \\
\left\langle S_j^{x} S_{j+1}^{z} S_{j+2}^{z} S_{j+3}^{x} \right\rangle 
& = \frac{1}{240}- \frac{1}{4}\ln 2+\frac{169}{240}\zeta(3) -\frac{5}{12} \ln 2 \cdot \zeta(3) 
-\frac{3}{20} \zeta(3)^2 -\frac{65}{96}\zeta(5)+\frac{5}{6} \ln 2 \cdot \zeta(5) \label{corr_n4}
\end{align}
Note that the other correlation functions among four lattice sites are 
expressed as a linear combination of the correlations 
above. For example we have 
\begin{align}
\left\langle S_j^{z} S_{j+1}^{z} S_{j+2}^{z} S_{j+3}^{z} \right\rangle 
&= \left\langle S_j^{x} S_{j+1}^{x} S_{j+2}^{z} S_{j+3}^{z} \right\rangle 
+\left\langle S_j^{x} S_{j+1}^{z} S_{j+2}^{x} S_{j+3}^{z} \right\rangle 
+\left\langle S_j^{x} S_{j+1}^{z} S_{j+2}^{z} S_{j+3}^{x} \right\rangle  \nonumber \\
&=\frac{1}{80}- \frac{1}{3}\ln 2+\frac{29}{30}\zeta(3) -\frac{2}{3} \ln 2 \cdot \zeta(3) 
-\frac{21}{80} \zeta(3)^2 -\frac{95}{96}\zeta(5)+\frac{35}{24} \ln 2 \cdot \zeta(5), \nonumber  \\
P(4) &= \frac{1}{16} + \frac{3}{4} \left\langle S_j^{z} S_{j+1}^{z} 
\right\rangle +\frac{1}{2} \left\langle S_j^{z} S_{j+2}^{z} \right\rangle 
+\frac{1}{4} \left\langle S_j^{z} S_{j+3}^{z} \right\rangle 
+\left\langle S_j^{z} S_{j+1}^{z} S_{j+2}^{z} S_{j+3}^{z} \right\rangle, 
\end{align}
etc....

In papers \cite{KSTS1,TKS,KSTS2} the above results were
generalized to the ${XXZ}$ Heisenberg chain both in massless and in massive
regime.  

Generalizing the results (\ref{corr_n4}) to the correlation functions 
on five lattice sites is a natural next problem. In principle, 
it is possible to calculate them similarly from the multiple integrals 
as was actually done for ${P(5)}$. It, however, will take 
a tremendous amount of work to complete them.  

It is worthy to note that all the above results have one specific 
feature, namely, they all are expressed in terms of one-dimensional
integrals. In series of papers \cite{bks1,bks2,bks3} this phenomenon 
was explained by means of the above mentioned connection of 
correlation functions in the inhomogeneous case with the quantum 
Knizhnik-Zamolodchikov equation, more precisely, by means of a duality 
between solutions to the qKZ equation on level -4 and 0. It was also 
argued that this fact has a deep mathematical origin based on theory of 
deformed hyper-elliptic integrals, 
symplectic group and special type of cohomologies investigated 
earlier by Nakayashiki and Smirnov \cite{ns,n}. 
One of basic outcomes of the above scheme is a general 
ansatz for correlation functions
in which only one transcendental function participates with some
rational functions as coefficients. Some special additional properties
derived from the qKZ equation allowed to fix $P(n)$ until $n=6$.

In recent paper \cite{BJMST} the properties of the qKZ equation have 
been further investigated and the above ansatz has been proved by means
of some special recursion relations which linearly connect correlation functions
at $n, n-1$ and $n-2$. Coefficients in these relations appeared to be 
related to special transfer matrices over an auxiliary space of 
``fractional dimension". The solution to these relations has also been
found in the paper \cite{BJMST} which gives an explicit result
for rational coefficients before the transcendental functions
staying in the ansatz for correlation functions.
In spite of the fact that this formula is explicit for any number of 
sites $n$, it is still hard to use it for getting manifest results for 
concrete $n$. 

In this paper we generalize the scheme suggested in paper \cite{bks1} 
to the case of an arbitrary correlation function. We use a set of 
relations for correlation functions that follow from the qKZ equation. 
We optionally call them ``first principle" relations. The claim is that
these relations together with the above mentioned ansatz completely
fix rational functions which participate in this ansatz. 
We find the whole set of correlation functions in case of $n=5$.
After taking homogeneous limit we come to the result in terms of
the Riemann zeta function in full agreement with the above conjecture, 
which now should be considered as proved. 
The ``first principle" relations as well as results for correlation functions
should be definitely equivalent to those given by the formula (3.22)
in the paper \cite{BJMST}. But it is still to show such an equivalence
explicitly.

\section{Quantum Knizhnik-Zamolodchikov 
equations and fundamental relations for correlation functions}
\setcounter{equation}{0}

In this section we would like to describe an algebraic
scheme which allows to calculate
any correlation function for the $XXX$ model.
In fact we generalize here the method applied in paper \cite{bks}
to the particular case of the EFP.

Let $R(\l)$ be the $R$-matrix for the $XXX$ model 
\be
R(\l)=
\frac{R_0(\l)}{\l+\pi i}
\left(
\begin{array}{cccc}
\l+\pi i&0&0&0\\
0&\l&\pi i&0\\
0&\pi i&\l&0\\
0&0&0&\l+\pi i
\end{array}
\right),
\label{R-m}
\ee
where
$$
R_0(\l)=-\frac{\Gamma\(\frac \l {2\pi i}\)
\Gamma\(\frac 1 2-\frac \l {2\pi i}\)}
{\Gamma\(-\frac \l {2\pi i}\)\Gamma\(\frac 1 2+\frac \l {2\pi i}\)}.
$$
Sometimes we shall also use notation
\be
\ov R(z)=
\left(
\begin{array}{cccc}
1&0&0&0\\
0&\frac{z}{z+i}&\frac{i}{z+i}&0\\
0&\frac{i}{z+i}&\frac{z}{z+i}&0\\
0&0&0&1
\end{array}
\right),
\label{R-m1}
\ee
where it is implied that $\l = \pi z$.

The corner stone of our scheme is the quantum Knizhnik-Zamolodchikov
equation [qKZ] \cite{KZ,FR,book,sm}. 
It is a linear system of equations for a vector function $g_n$. 
In particular, the qKZ equation on level $-4$ looks as follows
\beq
&g_n(\l _1,\cdots ,\l _{j+1},\l _j,\cdots,\l _{2n})_
{\e _1,\cdots ,\e _{j+1}',\e _j',\cdots,\e _{2n}}=&\label{symm}\\
&=R(\l _j -\l _{j+1})^{\e _{j},\e _{j+1}}_{\e _{j}',\e _{j+1}'}
\  \ g_n(\l _1,\cdots ,\l _{j},\l _{j+1},\cdots,\l _{2n})
_{\e _1,\cdots ,\e _{j},\e _{j+1},\cdots,\e _{2n}}
\nonumber
\eeq
\be
g_n(\l _1,\cdots ,\l _{2n-1},\l _{2n}+2\pi i)_
{\e _1,\cdots ,\e _{2n-1},\e _{2n}}
=
g_n(\l _{2n},\l _1,\cdots ,\l _{2n-1})
_{\e _{2n},\e _1,\cdots ,\e _{2n-1}}
\label{Rie}
\ee
It was suggested by the Kyoto school \cite{JMN} that
the correlation functions are connected with solutions to the qKZ 
equation. In particular, for the massless case Jimbo and Miwa \cite{JM}
found that one should take a special solution of the qKZ on level -4
which satisfies one additional constraint
\beq
&g_n(\l _1,\cdots ,\l _{j-1},\l _{j},\l _j-\pi i,\l _{j+2},\cdots,\l _{2n})_
{\e _1,\cdots ,\e _{j-1},\e _{j},\e _{j+1},\e _{j+2},\cdots,\e _{2n}}=&
\nonumber\\
&=\e _{j}\delta_{\e _{j},-\e _{j+1}}
\ g_{n-1}(\l _1,\cdots ,\l _{j-1},\l _{j+2},\cdots,\l _{2n})
_{\e _1,\cdots,\e _{j-1},\e _{j+2}, \cdots,\e _{2n}}
&\label{val1}
\eeq
Solutions to the above equations (\ref{Rie},\ref{val1}) are meromorphic functions 
with possible singularities at the points 
$$ \Im (\l _j-\l_k)=\pi  l, \quad l\in {\mbox{\bf Z}}\backslash 0$$
but regular at $ \Im (\l _j-\l_k)=\pm\pi $.

Due to Jimbo and Miwa, any correlation function 
$$
P_{\e_1,\ldots,\e_n}^{\e'_1,\ldots,\e'_n}\equiv 
\left\langle E_{\e_1}^{\e'_1}\ldots E_{\e_n}^{\e'_n}\right\rangle
$$
is given by the following formula
\be
P_{\e_1,\ldots,\e_n}^{\e'_1,\ldots,\e'_n}(z_1,\ldots,z_n)
= {g_n(\l_1,\ldots,\l_n,\l_n+\pi i,\ldots,\l_1+\pi i)}_
{\e_1,\ldots,\e_n,-\e'_n,\ldots,-\e'_1}
\label{Pgn}
\ee
where as above $\l_j=\pi  z_j$.

Let us list the ``first principle" relations for
the correlation functions which follow from
the qKZ equation:

\begin{itemize}

\item
Translational invariance

\be
P_{\e_1,\ldots,\e_n}^{\e'_1,\ldots,\e'_n}(z_1+x,\ldots,z_n+x)=
P_{\e_1,\ldots,\e_n}^{\e'_1,\ldots,\e'_n}(z_1,\ldots,z_n)
\label{transl}
\ee

\item
Transposition, Negating and Reverse order relations

\beq
&
P_{\e_1,\ldots,\e_n}^{\e'_1,\ldots,\e'_n}(z_1,\ldots,z_n)=
P_{\e'_1,\ldots,\e'_n}^{\e_1,\ldots,\e_n}(-z_1,\ldots,-z_n)=
&\nonumber\\
&P_{-\e_1,\ldots,-\e_n}^{-\e'_1,\ldots,-\e'_n}(z_1,\ldots,z_n)=
P_{\e_n,\ldots,\e_1}^{\e'_n,\ldots,\e'_1}(-z_n,\ldots,-z_1)&
\label{chain1}
\eeq

\item
Intertwining relation

\beq
&{\ov R}^{\e'_j\e'_{j+1}}_{\tilde\e'_j\tilde\e'_{j+1}}(z_j-z_{j+1})
P_{\ldots\e_{j+1},\e_j\ldots}^{\ldots\tilde\e'_{j+1},
\tilde\e'_j\ldots}(\ldots z_{j+1},z_j\ldots) =&\nonumber\\
&P_{\ldots\tilde\e_j,\tilde\e_{j+1}\ldots}^
{\ldots\e'_j,\e'_{j+1}\ldots}(\ldots z_j,z_{j+1}\ldots)
{\ov R}^{\tilde\e_j\tilde\e_{j+1}}_{\e_j\e_{j+1}}(z_j-z_{j+1})&
\label{intertwine}
\eeq

\item
First recurrent relation

\beq
&P_{\e_1,\e_2,\ldots,\e_n}^{\e'_1,\e'_2,\ldots,\e'_n}
(z+i,z,z_3\ldots,z_n)=
-\d_{\e_1,-\e_2}\e'_1\e_2
P_{-\e'_1,\e_3,\ldots,\e_n}^
{\e'_2,\e'_3,\ldots,\e'_n}(z,z_3\ldots,z_n)&\nonumber\\
&P_{\e_1,\e_2,\ldots,\e_n}^{\e'_1,\e'_2,\ldots,\e'_n}
(z-i,z,z_3\ldots,z_n)=
-\d_{\e'_1,-\e'_2}\e_1\e'_2
P_{\e_2,\e_3,\ldots,\e_n}^
{-\e_1,\e'_3,\ldots,\e'_n}(z,z_3\ldots,z_n)&
\label{chain2}
\eeq

\item
Second recurrent relation

\be
\lim_{z_j\rightarrow \infty}
P_{\e_1,\ldots,\e_j,\ldots,\e_n}^{\e'_1,\ldots,\e'_j,\ldots,\e'_n}
(z_1,\ldots,z_j,\ldots,z_n)=
\d_{\e_j,\e'_j}\;\frac{1}{2}\;
P_{\e_1,\ldots,\hat\e_j,\ldots,\e_n}^
{\e'_1,\ldots,\hat\e'_j,\ldots,\e'_n}
(z_1,\ldots,\hat z_j,\ldots,z_n)
\label{limit}
\ee

\item
Identity relation
$$
\;\sum_{
{\small
\begin{array}{c}
\e_1,\ldots,\e_n\\
\sum_i \e'_i = \sum_i \e_i 
\end{array}
}}
P_{\e_1,\ldots,\e_n}^{\e'_1,\ldots,\e'_n}(z_1,\ldots,z_n)
=
\;\sum_{
{\small
\begin{array}{c}
\e'_1,\ldots,\e'_n\\
\sum_i \e'_i = \sum_i \e_i 
\end{array}
}}
P_{\e_1,\ldots,\e_n}^{\e'_1,\ldots,\e'_n}(z_1,\ldots,z_n)=
$$
\be
=P_{+,\ldots,+}^{+,\ldots,+}(z_1,\ldots,z_n)
=P_{-,\ldots,-}^{-,\ldots,-}(z_1,\ldots,z_n)
\label{sum}
\ee

\item
Reduction relation

\beq
P_{+,\e_2,\ldots,\e_n}^
{+,\e'_2,\ldots,\e'_n}(z_1,z_2,\ldots,z_n) &+&
P_{-,\e_2,\ldots,\e_n}^
{-,\e'_2,\ldots,\e'_n}(z_1,z_2,\ldots,z_n)=
P_{\e_2,\ldots,\e_n}^
{\e'_2,\ldots,\e'_n}(z_2,\ldots,z_n)\nonumber\\
P_{\e_1,,\ldots,\e_{n-1},+}^
{\e'_1,\ldots,\e'_{n-1},+}(z_1,z_2,\ldots,z_n) &+&
P_{\e_1,,\ldots,\e_{n-1},-}^
{\e'_1,\ldots,\e'_{n-1},-}(z_1,z_2,\ldots,z_n)
=
P_{\e_1,\ldots,\e_{n-1}}^
{\e'_1,\ldots,\e'_{n-1}}(z_1,\ldots,z_{n-1})\nonumber\\
&&\label{chain3}
\eeq
\end{itemize}
The relations (\ref{transl}), (\ref{chain1}) and (\ref{chain3}) 
were discussed by Jimbo and Miwa in the paper \cite{JM}. 
The intertwining relation
(\ref{intertwine}) is the direct corollary of the qKZ equations.
The proof of the first and the second recurrent relations 
(\ref{chain2},\ref{limit})   can be performed in the same lines as
it was described in the paper \cite{bks}.  In particular, the relation
(\ref{limit}) was completely proved in the paper \cite{BJMST}.
The identity relation (\ref{sum}) is a direct consequence of the
$sl_2$ symmetry of the solution to the qKZ.

In papers \cite{bks1,bks2}, 
a general ansatz for the correlation functions
was suggested. Then it was proved 
\footnote{The function $\omega(\l)$ used in \cite{BJMST}
is related to the function $G$ by $\omega(\l)=G(i\l)+1/2$}
in \cite{BJMST}
\beq
&&P_{\e_1,\ldots,\e_n}^{\e'_1,\ldots,\e'_n}(z_1,\ldots,z_n)=
\nonumber\\
&&
\sum_{m=0}^{[\frac n 2 ]}
\sum_{1\le k_1\ne\cdots \ne k _{2m}\le n}
A_{\e_1,\ldots,\e_n}^{\e'_1,\ldots,\e'_n}
(k_1,\cdots ,k _{2m}|z_1,\ldots,z_n)
G(z_{k_1}-z_{k_2})\cdots G(z_{k_{2m-1}}-z_{k_{2m}})\nonumber\\
&&\quad 
\label{ansatz}
\eeq
where
\be
G(x)=2\sum_{k=1}^{\infty}(-1)^k\cdot k\cdot
{x^2+1\over x^2+k^2}
\label{G}
\ee
and $A_{\e_1,\ldots,\e_n}^{\e'_1,\ldots,\e'_n}
(k_1,\cdots ,k _{2m}|z_1,\ldots,z_n)$ are rational functions 
with known denominator and polynomial numerator of a known
degree in variables $z_1,\ldots,z_n$, namely,
\be
A_{\e_1,\ldots,\e_n}^{\e'_1,\ldots,\e'_n}
(k_1,\cdots ,k _{2m}|z_1,\ldots,z_n)=
\frac{Q_{\e_1,\ldots,\e_n}^{\e'_1,\ldots,\e'_n}
(k_1,\cdots ,k _{2m}|z_1,\ldots,z_n)}{\prod'_{i<j}(z_i-z_j)}
\label{Q}
\ee
where ${\prod'_{i<j}(z_i-z_j)}$ means the product in which
the differences with $\{i,j\}=\{k_{2l},k_{2l+1}\}$ and
$i,j\notin K$ should be excluded, $K$ is the set
$k_1,\cdots ,k _{2m}$.
$Q_{\e_1,\ldots,\e_n}^{\e'_1,\ldots,\e'_n}
(k_1,\cdots ,k _{2m}|z_1,\ldots,z_n)$ are polynomials of 
the variables $z$ which have the same degree in each 
variable as in the denominator and also the same
degree of homogeneity.  Some useful properties of the function ${G(x)}$ are listed below \cite{bks},
\begin{align}
G(-x) &= G(x), \\
G(\pm \infty) &=-\frac{1}{2}, \\
G(x - i) &=\alpha(x)+\gamma(x) G(x),
\end{align}
where 
\begin{align}
\alpha(x)=-\frac{x-2i}{x-i}, \ \ \ \ \gamma(x) = - \frac{x(x-2 i)}{x^2+1}.
\end{align}
Moreover ${G(x)}$ is a generating function of the alternating zeta functions 
\begin{align}
G(x) = -2(1+x^2) \sum_{k=0}^{\infty} (-1)^k x^{2 k} \zeta_a (2 k+1),
\end{align}
with 
\begin{align}
\zeta_a(s) \equiv \sum_{n=1}^{\infty} \frac{(-1)^{n-1}}{n^s} = (1- 2^{1-s}) \zeta(s).
\end{align}

Conjectural formula for the very first term when $m=0$ looks as follows
\be
A_{\e_1,\ldots,\e_n}^{\e'_1,\ldots,\e'_n}
(0|z_1,\ldots,z_n)=Q_{\e_1,\ldots,\e_n}^{\e'_1,\ldots,\e'_n}
(0|z_1,\ldots,z_n)=\frac{1}{(n+1)\binom{n}{r}}
\label{m=0}
\ee
where $r$ is the number of down (or up) spins in the set
$\e_1,\ldots,\e_n$ (or $\e'_1,\ldots,\e'_n$).

Our claim is that the ``first principle" relations (\ref{transl})-(\ref{chain3}) together 
with the ansatz (\ref{ansatz}) and conjecture (\ref{m=0}) completely defines all unknown 
coefficients in polynomials \\
$Q_{\e_1,\ldots,\e_n}^{\e'_1,\ldots,\e'_n}
(k_1,\cdots ,k _{2m}|z_1,\ldots,z_n)$ 
and hence the whole set of correlation functions. 

Here let us list the explicit form of the inhomogeneous correlaion functions  for ${n=1}$ and ${n=2}$.
\begin{align}
& P_{+}^{+}(z_1) = P_{-}^{-}(z_1)= \frac{1}{2}, \nonumber  \\
& P_{++}^{++}(z_1,z_2) =P_{--}^{--}(z_1,z_2) = \frac{1}{3} + \frac{1}{6}G(z_1-z_2), \nonumber \\
& P_{+-}^{+-}(z_1,z_2) =P_{-+}^{-+}(z_1,z_2) = \frac{1}{6} - \frac{1}{6}G(z_1-z_2), \nonumber \\ 
& P_{+-}^{-+}(z_1,z_2) =P_{-+}^{+-}(z_1,z_2) = \frac{1}{6} + \frac{1}{3}G(z_1-z_2).  \label{corrN2}
\end{align}
One can easily confirm these correlation functions satisfy the first principle relations.  

In the subsequent sections, we denote the inhomogeneous EFP as
\begin{align}
P_{n}(z_1,z_2,...,z_n) \equiv P_{++...,+}^{++...+}(z_1,z_2,...,z_n) \left( = P_{--...-}^{--...-}(z_1,z_2,...,z_n) \right). 
\end{align}
Especially we have ${P_2(z_1,z_2) = P_{++}^{++}(z_1,z_2)}$. The explicit form of the above inhomogeneous EFPs up 
to ${n=6}$ was obtained in \cite{bks}.  

\section{Inhomogeneous correlation functions for ${n=3}$}
\setcounter{equation}{0}
Here we consider the inhomogeneous correlation functions for ${n=3}$ and the corresponding first principle relations.
This case was considered as an example in the paper  \cite{BJMST}.
But we reproduce it here in order to make our approach more illustrative.
To describe the explicit expressions of ${P_{\e_1 \e_2 \e_3}^{\e'_1 \e'_2 \e'_3}(z_1,z_2,z_3)}$, we introduce some notations.  
\begin{align}
G_{ij} & \equiv G(z_{ij}), \ \ \ \ G_{123}  \equiv \frac{G_{12}}{z_{13}z_{23}} +\frac{G_{13}}{z_{12} z_{32}} +\frac{G_{23}}{z_{21} z_{31}}, 
\nonumber \\
G_{123}^{A} &=  \left(\frac{1}{z_{13}}-\frac{1}{z_{23}} \right) G_{12}+ \left(\frac{1}{z_{23}}-\frac{1}{z_{12}} \right) G_{13} 
+ \left(\frac{1}{z_{31}}-\frac{1}{z_{21}} \right) G_{23}, \label{G123A}
\end{align}
where we have also introduced an abbreviation ${z_{ij}=z_i-z_j}$. With these notations, the EFP ${P_3(z_1,z_2,z_3) \equiv 
P_{+++}^{+++}(z_1,z_2,z_3)}$ is given by  \cite{bks}
\begin{align}
P_3(z_1,z_2,z_3) &= \frac{1}{4} + \frac{1}{12} \left( G_{12}+ G_{13} + G_{23} \right) -\frac{1}{12} G_{123}.
\end{align}

Next we consider the inhomogeneous correlation functions ${P_{\e_1 \e_2 \e_3}^{\e'_1 \e'_2 \e'_3}(z_1,z_2,z_3)}$ 
in the sector ${\e_1+\e_2+\e_3=\e'_1+\e'_2+\e'_3=1}$.  For convenience, we first exhibit the final results in Table 1. 
There each row corresponds to the subscript ${ \e_1,\e_2,\e_3 }$, while each column 
corresponds to the superscript ${ \e'_1,\e'_2,\e'_3 }$. 

\begin{table}

\begin{tabular}{|c|c|c|c|} \hline 
${P_{\e_1 \e_2 \e_3}^{\e'_1 \e'_2 \e'_3}}$ 
& ${- + +}$ 
& ${+ - +}$ 
& ${+ + -}$  \\ \hline
${- + +}$ 
& 
\begin{tabular}{l}
${\frac{1}{12}+\frac{1}{12}(G_{23}-G_{12}-G_{13})}$ \\
${\hspace{2.5cm} + \frac{1}{12} G_{123}}$ 
\end{tabular}
& 
${\frac{1}{12} + \frac{1}{6}G_{12} + \frac{i}{12} G_{123}^{A}}$
&
\begin{tabular}{l}
${\frac{1}{12}+\frac{1}{6}G_{13}-\frac{1}{6} G_{123}}$ \\ 
${\hspace{2cm} - \frac{i}{12} G_{123}^{A}}$ 
\end{tabular}
\\ \hline
${+ - +}$ 
& ${\frac{1}{12} + \frac{1}{6}G_{12} - \frac{i}{12} G_{123}^{A}}$
& 
\begin{tabular}{l}
${\frac{1}{12} + \frac{1}{12} \left( G_{13}-G_{23}- G_{12} \right)}$ \\
${\hspace{2.5cm}  -\frac{1}{12} G_{123}}$
\end{tabular}
& ${\frac{1}{12} + \frac{1}{6}G_{23} + \frac{i}{12} G_{123}^{A}}$
\\ \hline
${+ + -}$ 
& 
\begin{tabular}{l}
${\frac{1}{12}+\frac{1}{6}G_{13}-\frac{1}{6} G_{123}}$ \\ 
${\hspace{2cm} + \frac{i}{12} G_{123}^{A}}$ 
\end{tabular} 
& 
${\frac{1}{12} + \frac{1}{6}G_{23} - \frac{i}{12} G_{123}^{A}}$
& 
\begin{tabular}{l}
${\frac{1}{12}+\frac{1}{12}(G_{12}-G_{13}-G_{23}) }$ \\ 
${\hspace{2.5cm} + \frac{1}{12} G_{123}}$ 
\end{tabular} \\ \hline
\end{tabular}
\caption{Inhomogeneous correlation functions for ${n=3}$ in the sector ${\e_1+\e_2+\e_3=1}$}
\end{table}


Below we show the strategy to calculate these correlation functions from our first principle relations. 
Assume we have already obtained the EFP, ${P_3(z_1,z_2,z_3)}$ and all the correlations for ${n=2}$.  
As is shown in Table 1, there are 9 correlation functions in the sector ${\e_1+\e_2+\e_3=\e'_1+\e'_2+\e'_3=1}$. 
However, we find, they are actually connected through the intertwining 
relation 
(\ref{intertwine}) as follows, 
\begin{align}
P_{+-+}^{-++}(z_1,z_2,z_3) &=\frac{z_{23}+i}{z_{23}} P_{++-}^{-++}(z_1,z_3,z_2) 
- \frac{i}{z_{23}} P_{++-}^{-++}(z_1,z_2,z_3), \nonumber  \\
P_{++-}^{+-+}(z_1,z_2,z_3) &=\frac{z_{21}+i}{z_{21}} P_{++-}^{-++}(z_2,z_1,z_3) 
- \frac{i}{z_{21}} P_{++-}^{-++}(z_1,z_2,z_3), \nonumber  \\
P_{-++}^{+-+}(z_1,z_2,z_3) &= P_{+-+}^{-++}(z_2,z_1,z_3) 
+ \frac{i}{z_{21}}\left\{ P_{-++}^{-++}(z_2,z_1,z_3) - P_{-++}^{-++}(z_1,z_2,z_3) \right\}, \nonumber  \\
P_{-++}^{++-}(z_1,z_2,z_3) &=\frac{z_{32}+i}{z_{32}} P_{-++}^{+-+}(z_1,z_3,z_2) 
- \frac{i}{z_{32}} P_{-++}^{+-+}(z_1,z_2,z_3), \nonumber  \\ 
P_{+-+}^{+-+}(z_1,z_2,z_3) &= P_{-++}^{-++}(z_2,z_1,z_3) + \frac{i}{z_{21}} 
\left\{ P_{+-+}^{-++}(z_2,z_1,z_3) - P_{+-+}^{-++}(z_1,z_2,z_3) \right\}, \nonumber \\
P_{+-+}^{++-}(z_1,z_2,z_3) &=\frac{z_{12}+i}{z_{12}} P_{-++}^{++-}(z_2,z_1,z_3) 
- \frac{i}{z_{12}} P_{-++}^{++-}(z_1,z_2,z_3), \nonumber \\
P_{++-}^{++-}(z_1,z_2,z_3) &= P_{+-+}^{+-+}(z_1,z_3,z_2) 
+ \frac{i}{z_{32}} \left\{ P_{++-}^{+-+}(z_1,z_3,z_2) - P_{++-}^{+-+}(z_1,z_2,z_3) \right\}.
\label{intertwine3}
\end{align}
These relations imply we need to calculate only two correlations, 
for example, ${P_{-++}^{-++}(z_1,z_2,z_3)}$ and ${P_{++-}^{-++}(z_1,z_2,z_3)}$. The remaining 
correlation functions are obtained immediately from (\ref{intertwine3}).  Moreover the diagonal one, 
${P_{-++}^{-++}(z_1,z_2,z_3)}$ can be derived easily from the reduction 
relation (\ref{chain3}), 
\begin{align}
P_{-++}^{-++}(z_1,z_2,z_3) &=P_2(z_2,z_3) - P_3(z_1,z_2,z_3) \nonumber  \\
&= \frac{1}{3}+\frac{1}{6} G_{23} -\left(\frac{1}{4}
+\frac{1}{12} \left( G_{12}+G_{13}+G_{23} \right) - \frac{1}{12} G_{123} \right) \nonumber  \\
&= \frac{1}{12}
+\frac{1}{12} \left( G_{23}-G_{12}-G_{13} \right) + \frac{1}{12} G_{123}.
\label{reduction3}
\end{align} 
Therefore our task reduces to calculate one off-diagonal correlation ${P_{++-}^{-++}(z_1,z_2,z_3)}$. 
Note that there are different ways to connect the correlations from (\ref{intertwine3}). 
They, however, are consistent with each other due to the Yang-Baxter relation for ${\ov R_{ij}(z)}$. 

Look at other first principle relations and apply them to ${P_{++-}^{-++}(z_1,z_2,z_3)}$.  
First, we find the reverse order relation gives 
\begin{align}
P_{++-}^{-++}(z_1,z_2,z_3) = P_{++-}^{-++}(z_3,z_2,z_1),   \label{n3eq0}
\end{align}
which means  ${P_{++-}^{-++}(z_1,z_2,z_3)}$ is symmetric with respect to the exchange 
${z_1 \leftrightarrow z_3}$. Next, the transposition relation ${P_{-++}^{++-}(z_1,z_2,z_3)=P_{++-}^{-++}(-z_1,-z_2,-z_3)}$ together with 
(\ref{intertwine3}), (\ref{reduction3}) and (\ref{n3eq0}) gives an equation 
\begin{align}
& P_{++-}^{-++}(-z_1,-z_2,-z_3) \nonumber  \\
& = \frac{z_{12}+i}{z_{12}} \frac{z_{32}+i}{z_{32}} P_{++-}^{-++}(z_1,z_2,z_3) 
-\frac{i}{z_{12}} \frac{z_{31}+i}{z_{31}} P_{++-}^{-++}(z_2,z_1,z_3) 
-\frac{i}{z_{32}} \frac{z_{13}+i}{z_{13}} P_{++-}^{-++}(z_1,z_3,z_2) \nonumber  \\
& - \frac{1}{z_{12} z_{32}} P_2(z_1,z_3) + \frac{i}{z_{31}} \frac{z_{32} +i}{z_{32}} P_2(z_1,z_2) 
+ \frac{i}{z_{13}} \frac{z_{12} +i}{z_{12}} P_2(z_2,z_3). \label{n3eq1}
\end{align}
The recurrent relations for ${P_{++-}^{-++}(z_1,z_2,z_3)}$ are given by 
\begin{align}
P_{++-}^{-++}(z+i,z,z_3) &= 0, \ \ 
P_{++-}^{-++}(z-i,z,z_3) = - P_{+-}^{-+}(z,z_3), \label{n3eq2} \\
\lim_{z_1 \to \infty} P_{++-}^{-++}(z_1,z_2,z_3) 
&= \lim_{z_3 \to \infty} P_{++-}^{-++}(z_1,z_2,z_3) =0, \nonumber  \\ 
\lim_{z_2 \to \infty} P_{++-}^{-++}(z_1,z_2,z_3) 
&= \frac{1}{2} P_{+-}^{-+}(z_1,z_3). \label{n3eq3}
\end{align} 
The identity relation gives a relation, 
\begin{align}
\frac{z_{32}+i}{z_{32}} P_{++-}^{-++}(z_1,z_2,z_3) + \frac{z_{23}+i}{z_{23}} P_{++-}^{-++}(z_1,z_3,z_2) 
= 2 P_3(z_1,z_2,z_3) - P_2(z_2,z_3). \label{n3eq4}
\end{align}
where we have used the relations (\ref{intertwine3}) and (\ref{reduction3}). 
Equations (\ref{n3eq0})--(\ref{n3eq4}) with the translational invariance 
\begin{align}
P_{++-}^{-++}(z_1+x,z_2+x,z_3+x)=P_{++-}^{-++}(z_1,z_2,z_3). \label{n3eq5}
\end{align}
are our first principle relations reduced to  ${P_{++-}^{-++}(z_1,z_2,z_3)}$.

According to the ansatz (\ref{ansatz}) and (\ref{m=0}), let us assume ${P_{++-}^{-++}(z_1,z_2,z_3)}$ 
in the form 
\begin{align}
P_{++-}^{-++}(z_1,z_2,z_3)&=\frac{1}{12} + \frac{c_{11}+c_{12}z_{13}+c_{13}z_{23}+c_{14}z_{13}z_{23}}{z_{13} z_{23}}G_{12}
\nonumber  \\ 
& \ \ \ \ + \frac{c_{21}+c_{22}z_{12}+c_{23}z_{32}+c_{24}z_{12}z_{32}}{z_{12} z_{32}}G_{13} \nonumber \\
& \ \ \ \ + \frac{c_{31}+c_{32}z_{21}+c_{33}z_{31}+c_{34}z_{21}z_{31}}{z_{21} z_{31}}G_{23},
\end{align}
where ${c_{ij}}$ are the coefficients which will be determined by the relations (\ref{n3eq0})--(\ref{n3eq4}). 
Note that from the symmetry relation (\ref{n3eq0}), we at once have several relations  
\begin{align}
c_{31}=c_{11}, \ \ c_{32}=c_{13}, \ \ c_{33}= c_{12}, \ \ c_{34}=c_{14}, \ \ c_{23}=c_{22}.
\end{align} 
Relations (\ref{n3eq1})--(\ref{n3eq4}) generate equations for the coefficients ${c_{ij}}$. 
Although they are an overdetermined system of the equations for ${c_{ij}}$, one can figure 
out it has a unique solution  
\begin{eqnarray}
& c_{11} =c_{21}= c_{31} = -\frac{1}{6}, \ \ \ \ c_{12}=-c_{13}=c_{22}=c_{23}=-c_{32}=c_{33} = -\frac{i}{12}, \nonumber  \\
& c_{14} =c_{34}=0, \ \ \ \ c_{24} =\frac{1}{6}.
\end{eqnarray} 
Namely the correlation function ${P_{++-}^{-++}(z_1,z_2,z_3)}$ is determined as 
\begin{align}
P_{++-}^{-++}(z_1,z_2,z_3) &= \frac{1}{12} + 
\frac{1}{6} \left\{- \frac{1}{z_{13} z_{23}} + \frac{i}{2} \left( -\frac{1}{z_{23}} + \frac{1}{z_{13}} \right)  \right\} G_{12} \nonumber  \\
&+\frac{1}{6} \left\{ 1-  \frac{1}{z_{12} z_{32}} - \frac{i}{2} \left( \frac{1}{z_{32}} + \frac{1}{z_{12}} \right)  \right\} G_{13} \nonumber  \\
& +\frac{1}{6} \left\{ -  \frac{1}{z_{21} z_{31}} + \frac{i}{2} \left( \frac{1}{z_{31}} - \frac{1}{z_{21}} \right)  \right\} G_{23} \nonumber  \\
&= \frac{1}{12} + \frac{1}{6}G_{13} - \frac{1}{6} G_{123} + \frac{i}{12} G^{A}_{123}. 
\end{align}

Other correlation functions are obtained from the intertwining relations (\ref{intertwine3}). 
In this way we could calculate all the inhomogeneous correlation functions as shown in Table 1. The correlation 
functions in the sector ${\e_1+\e_2+\e_3=\e'_1+\e'_2+\e'_3=-1}$ are simply obtained by the negation 
relations, 
\begin{align}
P_{\e_1 \e_2 \e_3}^{\e'_1 \e'_2 \e'_3}(z_1,z_2,z_3) 
= P_{-\e_1, -\e_2,-\e_3}^{-\e'_1, -\e'_2, -\e'_3}(z_1,z_2,z_3).
\end{align}
Now let us think of an inhomogeneous generalization of the next-nearest neighbor correlation function,
\begin{align}
\left\langle S_1^{z} S_3^{z} \right\rangle (z_1,z_2,z_3) 
&= \frac{1}{2} \left\{ P_3(z_1,z_2,z_3)-P_{++-}^{++-}(z_1,z_2,z_3) 
+P_{+-+}^{+-+}(z_1,z_2,z_3)- P_{-++}^{-++}(z_1,z_2,z_3) \right\} \nonumber  \\
&=\frac{1}{4} -P_2(z_1,z_2)-P_2(z_2,z_3)+2 P_3(z_1,z_2,z_3) 
=\frac{1}{12} + \frac{1}{6} G_{13} - \frac{1}{6} G_{123} \nonumber  \\
&= \frac{1}{12} + \frac{1}{6} G(z_{13}) - \frac{1}{6} \left\{ \frac{G(z_{12})}{z_{13} z_{23}} 
+ \frac{G(z_{13})}{z_{12} z_{32}} + \frac{G(z_{23})}{z_{21} z_{31}} \right\}.
\label{inhomogeneous_next_nearest}
\end{align}
Expanding ${G}$-function as
\begin{equation}
G(z) = -2(1+z^2) \left\{ \zeta_a(1) - z^2 \zeta_a(3)+ \cdots \right\},
\end{equation} 
and substituting it into (\ref{inhomogeneous_next_nearest}), one can take the homogeneous limit ${z_1,z_2,z_3 \to 0}$,
\begin{align}
\lim_{z_i \to 0} \left\langle S_1^{z} S_3^{z} \right\rangle (z_1,z_2,z_3) 
&= \frac{1}{12} - \frac{1}{3} \zeta_a(1) - \left(\zeta_a (1) - \zeta_a (3) \right) \nonumber  \\
&= \frac{1}{12} -\frac{4}{3} \ln 2  +\frac{3}{4} \zeta(3),
\label{hom_limit_next_nearest}
\end{align}
which reproduces the expression (\ref{2nd})

\section{Inhomogeneous correlation functions for ${n=4}$}
\setcounter{equation}{0}
Next we consider the inhomogeneous correlation functions for ${n=4}$,
${P_{\e_1 \e_2 \e_3 \e_4}^{\e'_1 \e'_2 \e'_3 \e'_4}(z_1,z_2,z_3,z_4)}$.  \\ 
The EFP ${P_4(z_1,z_2,z_3,z_4)= P_{++++}^{++++}(z_1,z_2,z_3,z_4)}$ was calculated in \cite{bks}, 
\begin{align}
P_4(z_1,z_2,z_3,z_4)=& \frac{1}{5} +A_{4,1}^{(0)}(z_1,z_2,z_3,z_4) G(z_{12}) +A_{4,1}^{(0)}(z_1,z_3,z_2,z_4) G(z_{13})   \nonumber  \\
& + A_{4,1}^{(0)}(z_1,z_4,z_3,z_2) G(z_{14}) + A_{4,1}^{(0)}(z_3,z_2,z_1,z_4) G(z_{23}) \nonumber  \\
& + A_{4,1}^{(0)}(z_4,z_2,z_3,z_1) G(z_{24}) + A_{4,1}^{(0)}(z_4,z_3,z_2,z_1) G(z_{34}) \nonumber  \\
& + A_{4,2}^{(0)}(z_1,z_2,z_3,z_4) G(z_{12}) G(z_{34}) + A_{4,2}^{(0)}(z_1,z_3,z_2,z_4) G(z_{13}) G(z_{24}) \nonumber  \\
& + A_{4,2}^{(0)}(z_1,z_4,z_3,z_2) G(z_{14}) G(z_{23}), \nonumber  
\end{align}
with
\begin{align}
A_{4,i}^{(0)}(z_1,z_2,z_3,z_4) &= \frac{Q_{4,i}^{(0)}(z_1,z_2,z_3,z_4)}{z_{13}z_{14} z_{23} z_{24}}, \ \ \ \ (i=1,2),  \nonumber  \\
Q_{4,2}^{(0)}(z_1,z_2,z_3,z_4) &= \frac{1}{36} \left\{ (z_{13} z_{24}-1)(z_{14} z_{23}-1)+\frac{2}{5}\left(z_{12}^2+\frac{3}{2} \right) 
\left(z_{34}^2+\frac{3}{2}\right) + \frac{3}{2} \right\}, \nonumber \\
Q_{4,1}^{(0)}(z_1,z_2,z_3,z_4) &= 2 Q_{4,2}^{(0)}(z_1,z_2,z_3,z_4)- \frac{1}{60}(z_{12}^2+4)(z_{34}^2+1). \label{corrP4}
\end{align}
Let us think of the correlation functions in the sector 
${\e_1+\e_2+\e_3+\e_4 = \e_1'+\e_2'+\e_3'+\e_4'=2}$. 
There are 16 correlations in this sector. 
They, however, are connected via the intertwining relations similarly in the case of ${n=3}$. 
In fact, if one can calculate a single correlation, for example, ${P_{+++-}^{-+++}(z_1,z_2,z_3,z_4)}$, other correlation functions 
are obtained with the help of the intertwining relations and the reduction relation 
\begin{align}
P_{-+++}^{-+++}(z_1,z_2,z_3,z_4) = P_3(z_2,z_3,z_4)- P_4(z_1,z_2,z_3,z_4).
\end{align}  
Therefore we can invoke the same strategy as was done for the calculation of 
${P_{++-}^{-++}(z_1,z_2,z_3)}$ in the previous section. Namely, we reduce the first principle 
relations to those for a single functions ${P_{+++-}^{-+++}(z_1,z_2,z_3,z_4)}$ and solve them. 
Below we list the equations corresponding to (\ref{n3eq0})--(\ref{n3eq4}) for ${n=3}$. 
\begin{itemize}
\item symmetry relations 
\begin{align}
P_{+++-}^{-+++}(z_1,z_2,z_3,z_4)=P_{+++-}^{-+++}(z_1,z_3,z_2,z_4)=P_{+++-}^{-+++}(z_4,z_3,z_2,z_1)
\label{n4eq0}
\end{align}
\item transposition relation
\begin{align}
& P_{+++-}^{-+++}(-z_1,-z_2,-z_3,-z_4) \nonumber  \\
 & = \frac{(z_{12}+i)(z_{13}+i)(z_{42}+i)(z_{43}+i)}{z_{12} z_{13} z_{42} z_{43}} 
P_{+++-}^{-+++}(z_1,z_2,z_3,z_4) 
\nonumber  \\
& - \frac{i(z_{12}+i) (z_{14}+i) (z_{32}+i) }{z_{12} z_{14} z_{32} z_{43}}
P_{+++-}^{-+++}(z_1,z_2,z_4,z_3) \nonumber  \\
& - \frac{i(z_{13}+i) (z_{14}+i) (z_{23}+i) }{z_{13} z_{14} z_{23} z_{42}}
P_{+++-}^{-+++}(z_1,z_3,z_4,z_2) \nonumber  \\
& - \frac{i(z_{43}+i) (z_{41}+i) (z_{23}+i) }{z_{43} z_{41} z_{23} z_{12}}
P_{+++-}^{-+++}(z_2,z_1,z_3,z_4) \nonumber \\
& 
- \frac{i(z_{42}+i) (z_{41}+i) (z_{32}+i) }{z_{42} z_{41} z_{32} z_{13}}
P_{+++-}^{-+++}(z_3,z_1,z_2,z_4) \nonumber \\ 
& + \frac{1+i \left( z_{12}+z_{43} \right)+ z_{12} z_{34} +z_{13} z_{24}}{z_{12} z_{13} z_{24} z_{34}}
P_{+++-}^{-+++}(z_3,z_1,z_2,z_4) \nonumber  \\
& + \frac{i(z_{42}+i) (z_{43}+i)}{z_{42} z_{43} z_{41}} P_3(z_1,z_2,z_3) 
  + \frac{i(z_{12}+i) (z_{13}+i)}{z_{12} z_{13} z_{14}} P_3(z_2,z_3,z_4) 
  + \frac{z_{32}+i}{z_{32} z_{43} z_{31}} P_3(z_1,z_2,z_4) \nonumber \\
&  + \frac{z_{23}+i}{z_{23} z_{42} z_{21}} P_3(z_2,z_3,z_4)
\label{n4eq1}
\end{align}
\item
first recurrent relations 
\begin{align}
\lim_{z_1\to \infty} P_{+++-}^{-+++}(z_1,z_2,z_3,z_4)  &=0, \nonumber  \\
\lim_{z_2 \to \infty} P_{+++-}^{-+++}(z_1,z_2,z_3,z_4) &=\frac{1}{2} P_{++-}^{-++}(z_1,z_3,z_4) 
\label{n4eq2}
\end{align}
\item
second recurrent relation 
\begin{align}
P_{+++-}^{-+++}(z+i,z,z_3,z_4) &=0, \nonumber \\
P_{+++-}^{-+++}(z-i,z,z_3,z_4) &= - P_{++-}^{-++}(z,z_3,z_4)
\label{n4eq3}
\end{align}
\item
identity relation 
\begin{align}
& \frac{z_{42}+i}{z_{42}} \frac{z_{43}+i}{z_{43}} P_{+++-}^{-+++}(z_1,z_2,z_3,z_4) 
+\frac{z_{23}+i}{z_{23}} \frac{z_{24}+i}{z_{24}} P_{+++-}^{-+++}(z_1,z_3,z_4,z_2) \nonumber  \\
&+\frac{z_{32}+i}{z_{32}} \frac{z_{34}+i}{z_{34}} P_{+++-}^{-+++}(z_1,z_2,z_4,z_3) 
= 2 P_4(z_1,z_2,z_3,z_4) - P_3(z_2,z_3,z_4)
\label{n4eq4}
\end{align}
\item 
translational invariance
\begin{align}
P_{+++-}^{-+++}(z_1+x,z_2+x,z_3+x,z_4+x) = P_{+++-}^{-+++}(z_1,z_2,z_3,z_4)
\label{n4eq5}
\end{align}
\end{itemize}
According to the ansatz (\ref{ansatz}), (\ref{m=0}), 
${P_{+++-}^{-+++}(z_1,z_2,z_3,z_4)}$ may be written in the form 
\begin{align}
P_{+++-}^{-+++}(z_1,z_2,z_3,z_4)
& =\frac{1}{20} 
  + A_{4,1}^{(1)}(z_1,z_2,z_3,z_4) G(z_{12}) + A_{4,1}^{(1)}(z_1,z_3,z_2,z_4) G(z_{13}) \nonumber  \\ 
& + A_{4,2}^{(1)}(z_1,z_4,z_2,z_3) G(z_{14}) + A_{4,3}^{(1)}(z_3,z_2,z_1,z_4) G(z_{23})  \nonumber  \\
& + A_{4,1}^{(1)}(z_4,z_2,z_3,z_1) G(z_{24}) + A_{4,1}^{(1)}(z_4,z_3,z_2,z_1) G(z_{34}) \nonumber  \\
& + A_{4,4}^{(1)}(z_1,z_2,z_3,z_4) G(z_{12}) G(z_{34}) 
  + A_{4,4}^{(1)}(z_1,z_3,z_2,z_4) G(z_{13}) G(z_{24}) \nonumber  \\
& + A_{4,5}^{(1)}(z_1,z_4,z_3,z_2) G(z_{14}) G(z_{23}), \nonumber  \\
A_{4,i}^{(1)}(z_1,z_2,z_3,z_4) 
&= \frac{Q_{4,i}^{(1)}(z_1,z_2,z_3,z_4)}{z_{13} z_{14} z_{23} z_{24}}, \ \ \ \  \ \ \ \ (i=1,...,5),
\label{n4assume}
\end{align}
where we have taken the symmetry relations (\ref{n4eq0}) into account. 
We can assume the polynomials ${Q_{4,i}^{(1)}(z_1,z_2,z_3,z_4)}$ as 
\begin{align}
Q_{4,i}^{(1)}(z_1,z_2,z_3,z_4) = \sum_{0 \leq j_1,j_2,j_3,j_4 \leq 2 \atop j_1+j_2+j_3+j_4 \leq 4} C_{4,i}(j_1,j_2,j_3,j_4) 
z_1^{j_1} z_2^{j_2} z_3^{j_3} z_4^{j_4}.
\label{n4assumeQ} 
\end{align}
Substituting (\ref{n4assume}) and (\ref{n4assumeQ}) into (\ref{n4eq1}),...,(\ref{n4eq5}), we can extract the 
equations for coefficients ${C_{4,i}(j_1,j_2,j_3,j_4)}$ and solve them. Actually we have made 
a {\it Mathematica} program to perform the calculation. The result is 
\begin{align}
& Q_{4,1}^{(1)}(z_1,z_2,z_3,z_4) \nonumber  \\
&=\frac{19}{90} + \frac{7}{80} z_{12}^2-\frac{1}{12} z_{13}^2 - \frac{1}{144} z_{14}^2 -\frac{1}{36} z_{23}^2 
- \frac{1}{48} z_{24}^2 + \frac{1}{30} z_{34}^2 - \frac{1}{144}z_{13}^2 z_{24}^2 
+ \frac{1}{144} z_{14}^2 z_{23}^2 + \frac{1}{720} z_{12}^2 z_{34}^2 \nonumber  \\  
& + i \bigg( \frac{1}{9} z_{13} - \frac{1}{36} z_{24} + \frac{1}{72} z_{12}^2 z_{34} - \frac{1}{18} z_{12} z_{34}^2 
-\frac{1}{36} z_{13}^2 z_{24} + \frac{5}{72} z_{13} z_{24}^2 - \frac{1}{24} z_{14}^2 z_{23} 
\bigg), \nonumber  \\
& Q_{4,2}^{(1)}(z_1,z_2,z_3,z_4) \nonumber  \\
&= \frac{19}{90} + \frac{59}{360} z_{12}^2-\frac{1}{12} z_{13}^2-\frac{1}{12} z_{23}^2 
-\frac{1}{12} z_{14}^2 - \frac{1}{12} z_{24}^2 + \frac{4}{45} z_{34}^2 + \frac{1}{18}z_{13}^2 z_{24}^2 
+ \frac{1}{18} z_{14}^2 z_{23}^2 - \frac{17}{360} z_{12}^2 z_{34}^2 \nonumber  \\  
& + i \bigg( \frac{1}{9} z_{13} + \frac{1}{9} z_{24} - \frac{1}{18} z_{13}^2 z_{24} - \frac{1}{18} z_{13} z_{24}^2 
- \frac{1}{18} z_{14}^2 z_{23} -\frac{1}{18} z_{14} z_{23}^2 \bigg), \nonumber  \\
& Q_{4,3}^{(1)}(z_1,z_2,z_3,z_4) \nonumber  \\
&= \frac{19}{90} + \frac{29}{360} z_{12}^2-\frac{1}{48} z_{13}^2-\frac{1}{48} z_{23}^2 
-\frac{1}{48} z_{14}^2 - \frac{1}{48} z_{24}^2 + \frac{17}{360} z_{34}^2 + \frac{1}{72}z_{13}^2 z_{24}^2 
+ \frac{1}{72} z_{14}^2 z_{23}^2 - \frac{1}{180} z_{12}^2 z_{34}^2 \nonumber  \\  
& + i \bigg( -\frac{1}{36} z_{13} - \frac{1}{36} z_{24} + \frac{1}{72} z_{13}^2 z_{24} + \frac{1}{72} z_{13} z_{24}^2 
+ \frac{1}{72} z_{14}^2 z_{23} +\frac{1}{72} z_{14} z_{23}^2 \bigg), \nonumber 
\end{align}
\begin{align}
& Q_{4,4}^{(1)}(z_1,z_2,z_3,z_4) \nonumber  \\
&= \frac{7}{45} + \frac{1}{15} z_{12}^2-\frac{1}{24} z_{13}^2-\frac{1}{72} z_{14}^2 -\frac{1}{72} z_{23}^2 
- \frac{1}{24} z_{24}^2 + \frac{1}{15} z_{34}^2 + \frac{1}{360}z_{12}^2 z_{34}^2 
- \frac{1}{72} z_{13}^2 z_{24}^2 + \frac{1}{72} z_{14}^2 z_{23}^2 \nonumber  \\  
& + i \bigg( \frac{1}{18} z_{13} - \frac{1}{18} z_{24} + \frac{1}{36} z_{12}^2 z_{34} -\frac{1}{36} z_{12} z_{34}^2 
- \frac{1}{18} z_{13}^2 z_{24} + \frac{1}{18} z_{13} z_{24}^2 \bigg), \nonumber \\
& Q_{4,5}^{(1)}(z_1,z_2,z_3,z_4) \nonumber  \\
&= \frac{7}{45} + \frac{17}{180} z_{12}^2-\frac{1}{24} z_{13}^2-\frac{1}{24} z_{14}^2 -\frac{1}{24} z_{23}^2 
- \frac{1}{24} z_{24}^2 + \frac{17}{180} z_{34}^2 - \frac{1}{90}z_{12}^2 z_{34}^2 
+ \frac{1}{36} z_{13}^2 z_{24}^2 + \frac{1}{36} z_{14}^2 z_{23}^2 \nonumber  \\  
& + i \bigg( \frac{1}{18} z_{13} + \frac{1}{18} z_{24} - \frac{1}{36} z_{13}^2 z_{24} -\frac{1}{36} z_{13} z_{24}^2 
- \frac{1}{36} z_{14}^2 z_{23} - \frac{1}{36} z_{14} z_{23}^2 \bigg). \label{corrN4sec1}
\end{align}

\begin{table}
\begin{center}
\small
\begin{tabular}{|c|c|c|c|c|} \hline 
${P_{\e_1 \e_2 \e_3 \e_4}^{\e_1' \e_2' \e_3' \e_4'}}$ & ${- + + +}$ & ${+ - + +}$ & ${+ + - +}$ & ${+ + + -}$ \\ \hline
${- + + +}$ & 
\begin{tabular}{l}
${P_3(z_2,z_3,z_4)}$ \\ 
${- P_4(z_1,z_2,z_3,z_4)}$ 
\end{tabular}& 
\begin{tabular}{l}
${A(z_1,z_2,z_3,z_4)}$ \\ 
${- i X(z_1,z_2,z_3,z_4)}$ 
\end{tabular}& 
\begin{tabular}{l}
${B(z_1,z_2,z_3,z_4)}$ \\ 
${- i Y(z_1,z_2,z_3,z_4)}$ 
\end{tabular}&
\begin{tabular}{l}
${C(z_1,z_2,z_3,z_4)}$ \\ 
${- i Z(z_1,z_2,z_3,z_4)}$ 
\end{tabular}\\ \hline
${+ - + +}$ & 
\begin{tabular}{l}
${A(z_1,z_2,z_3,z_4)}$ \\ 
${+ i X(z_1,z_2,z_3,z_4)}$ 
\end{tabular}& 
\begin{tabular}{l}
${D(z_1,z_2,z_3,z_4)}$ \\ 
\end{tabular}& 
\begin{tabular}{l}
${E(z_1,z_2,z_3,z_4)}$ \\ 
${+ \frac{i}{24}\left(G^{A}_{123}+G^{A}_{234} \right)}$ 
\end{tabular}&
\begin{tabular}{l}
${B(z_4,z_3,z_2,z_1)}$ \\ 
${- i Y(z_4,z_3,z_2,z_1)}$ 
\end{tabular}\\ \hline
${+ + - +}$ & 
\begin{tabular}{l}
${B(z_1,z_2,z_3,z_4)}$ \\ 
${+ i Y(z_1,z_2,z_3,z_4)}$ 
\end{tabular}& 
\begin{tabular}{l}
${E(z_1,z_2,z_3,z_4)}$ \\ 
${- \frac{i}{24}\left(G^{A}_{123}+G^{A}_{234} \right)}$ 
\end{tabular}& 
\begin{tabular}{l}
${D(z_4,z_3,z_2,z_1)}$ \\ 

\end{tabular}&
\begin{tabular}{l}
${A(z_4,z_3,z_2,z_1)}$ \\ 
${- i X(z_4,z_3,z_2,z_1)}$ 
\end{tabular}\\ \hline
${+ + + -}$ & 
\begin{tabular}{l}
${C(z_1,z_2,z_3,z_4)}$ \\ 
${+ i Z(z_1,z_2,z_3,z_4)}$ 
\end{tabular}& 
\begin{tabular}{l}
${B(z_4,z_3,z_2,z_1)}$ \\ 
${+ i Y(z_4,z_3,z_2,z_1)}$ 
\end{tabular}& 
\begin{tabular}{l}
${A(z_4,z_3,z_2,z_1)}$ \\ 
${+ i X(z_4,z_3,z_2,z_1)}$ 
\end{tabular}&
\begin{tabular}{l}
${P_3(z_1,z_2,z_3)}$ \\ 
${- P_4(z_1,z_2,z_3,z_4)}$ 
\end{tabular}\\ \hline
\end{tabular}
\end{center}
\caption{Inhomogeneous correlation functions for ${n=4}$ in the sector ${\e_1+\e_2+\e_3+\e_4=2}$}
\end{table}

Unfortunately we could not find a further compact way to express the results above. Other correlation functions in this sector 
are calculated by the intertwining relations.  As it is not possible to present all the explicit results here, we instead show the formal 
expressions in Table 2. Here ${C(z_1,z_2,z_3,z_4)}$ and ${Z(z_1,z_2,z_3,z_4)}$ are respectively the real and the 
imaginary part of the correlation function ${P_{+++-}^{-+++}(z_1,z_2,z_3,z_4)}$ (\ref{corrN4sec1}). Similarly, 
${A(z_1,z_2,z_3,z_4)}$, ${B(z_1,z_2,z_3,z_4)}$, ${D(z_1,z_2,z_3,z_4)}$, ${E(z_1,z_2,z_3,z_4)}$ and ${X(z_1,z_2,z_3,z_4)}$, 
${Y(z_1,z_2,z_3,z_4)}$ are respectively the real and the imaginary parts of other correlation functions. Using the intertwining 
relations, they are successively calculated from ${C(z_1,z_2,z_3,z_4)}$ and ${Z(z_1,z_2,z_3,z_4)}$, 
\begin{align}
B(z_1,z_2,z_3,z_4) &= C(z_1,z_2,z_4,z_3) + \frac{1}{z_{34}} \left\{ Z(z_1,z_2,z_3,z_4)-Z(z_1,z_2,z_4,z_3)  \right\}, \nonumber \\ 
Y(z_1,z_2,z_3,z_4) &= Z(z_1,z_2,z_4,z_3) + \frac{1}{z_{34}} \left\{ C(z_1,z_2,z_4,z_3)-C(z_1,z_2,z_3,z_4)  \right\}, \nonumber \\ 
A(z_1,z_2,z_3,z_4) &= B(z_1,z_3,z_2,z_4) + \frac{1}{z_{23}} \left\{ Y(z_1,z_2,z_3,z_4)-Y(z_1,z_3,z_2,z_4)  \right\}, \nonumber \\ 
X(z_1,z_2,z_3,z_4) &= Y(z_1,z_3,z_2,z_4) + \frac{1}{z_{23}} \left\{ B(z_1,z_3,z_2,z_4)-B(z_1,z_2,z_3,z_4)  \right\}, \nonumber \\ 
D(z_1,z_2,z_3,z_4) &= P_3(z_1,z_3,z_4) -P_4(z_2,z_1,z_3,z_4) + \frac{1}{z_{21}} \left\{ X(z_1,z_2,z_3,z_4)- X(z_2,z_1,z_3,z_4) \right\}, \nonumber  \\
E(z_1,z_2,z_3,z_4) &= B(z_2,z_1,z_3,z_4) + \frac{1}{z_{21}} \left\{ Y(z_1,z_2,z_3,z_4) - Y(z_2,z_1,z_3,z_4) \right\}.
\end{align}
Note that the diagonal correlation functions ${P_{+-++}^{+-++}(z_1,z_2,z_3,z_4)}$ and ${P_{++-+}^{++-+}(z_1,z_2,z_3,z_4)}$ have no imaginary parts. 
Note also that we have found the imaginary part of ${P_{++-+}^{+-++}(z_1,z_2,z_3,z_4)}$ is simply expressed in terms of the function 
${G_{123}^A}$ introduced in (\ref{G123A}).  These properties are reflected in some symmetry relations of the functions ${A(z_1,z_2,z_3,z_4)}$,...,
${Z(z_1,z_2,z_3,z_4)}$. For example we have 
\begin{align}
A(z_1,z_2,z_3,z_4) &= A(z_2,z_1,z_3,z_4), \ \ E(z_1,z_2,z_3,z_4) = E(z_1,z_3,z_2,z_4), \nonumber \\
Y(z_1,z_2,z_3,z_4) &= Y(z_2,z_1,z_3,z_4) + \frac{1}{24} \left( G_{123}^A +G_{234}^A-G_{213}^A-G_{134}^A \right).
\label{prop1}
\end{align}
We also remark the following relations hold from the identity relations. 
\begin{align}
& A(z_1,z_2,z_3,z_4)+ B(z_1,z_1,z_3,z_4)+ C(z_1,z_2,z_3,z_4)= 2 P_4(z_1,z_2,z_3,z_4) - P_3(z_2,z_3,z_4), \nonumber  \\
& A(z_1,z_2,z_3,z_4)+D(z_1,z_2,z_3,z_4)+E(z_1,z_2,z_3,z_4)+B(z_4,z_3,z_2,z_1)= P_4(z_1,z_2,z_3,z_4), \nonumber \\
& X(z_1,z_2,z_3,z_4)-Y(z_4,z_3,z_2,z_1)-\frac{1}{24}(G_{123}^A+G_{234}^A)=0, \nonumber  \\
& X(z_1,z_2,z_3,z_4)+Y(z_1,z_2,z_3,z_4)+Z(z_1,z_2,z_3,z_4)=0. \label{prop2}
\end{align}
Unfortunately we have not been able to afford a nicer compact way to describe the functions ${A(z_1,z_2,z_3,z_4)}$,...,
${Z(z_1,z_2,z_3,z_4)}$, from which the relations such as (\ref{prop1}), (\ref{prop2}) follow naturally. 

Next let us consider the sector ${\e_1+\e_2+\e_3+\e_4 = \e_1'+\e_2'+\e_3'+\e_4'=0}$. There are 36 correlation functions in this sector. But 
according to (\ref{chain1}), we need to know only 12 correlation functions ${C_1(z1_,z_2,z_3,z_4)}$,...,${C_{12}(z_1,z_2,z_3,z_4)}$ as shown 
in Table 3.

\begin{table}
\begin{center}
\small
\begin{tabular}{|c|c|c|c|c|c|c|} \hline 
${P_{\e_1 \e_2 \e_3 \e_4}^{\e_1' \e_2' \e_3' \e_4'}}$ & ${- - + +}$ & ${- + - +}$ & ${- + + -}$ & ${+ - - +}$ & ${+ - + - }$ & ${+ + - - }$\\ \hline
${- - + +}$ 
& ${C_1(\{z_i\})}$
& ${C_2(\{-z_i\})}$
& ${C_3(\{-z_i\})}$
& ${C_4(\{-z_i\})}$
& ${C_5(\{-z_i\})}$
& ${C_6(\{z_i\})}$
\\ \hline
${- + - +}$ 
& ${C_2(\{z_i\})}$
& ${C_7(\{z_i\})}$
& ${C_8(\{-z_i\})}$
& ${C_9(\{-z_i\})}$
& ${C_{10}(\{z_i\})}$
& ${C_5 (\{z_{i}\})}$
\\ \hline
${- + + -}$ 
& ${C_3(\{z_i \})}$
& ${C_8(\{z_i \})}$
& ${C_{11}(\{z_i \})}$
& ${C_{12}(\{z_i \})}$
& ${C_9 (\{z_i \})}$
& ${C_4 (\{z_i \})}$
\\ \hline
${+ - - +}$ 
& ${C_4(\{z_i\})}$
& ${C_9(\{z_i\})}$
& ${C_{12}(\{z_i \})}$
& ${C_{11}(\{z_i \})}$
& ${C_8(\{z_i \})}$
& ${C_3(\{z_i \})}$
\\ \hline
${+ - + -}$ 
& ${C_5(\{z_i\})}$
& ${C_{10}(\{z_i\})}$
& ${C_9(\{-z_i \})}$
& ${C_8(\{-z_i \})}$
& ${C_7(\{z_i \})}$
& ${C_2(\{z_i \})}$
\\ \hline
${+ + - -}$ 
& ${C_6(\{z_i \})}$
& ${C_5(\{-z_i \})}$
& ${C_4(\{-z_i \})}$
& ${C_3(\{-z_i \})}$
& ${C_2(\{-z_i \})}$
& ${C_1(\{z_i \})}$
\\ \hline
\end{tabular}
\end{center}
\caption{Inhomogeneous correlation functions for ${n=4}$ in the sector ${\e_1+\e_2+\e_3+\e_4=0}$}
\end{table}
In the present case, they can be derived from those we have obtained already as follows.  
\begin{align}
C_1(z_1,z_2,z_3,z_4) &= P_{-++}^{-++}(z_2,z_3,z_4)-P_{+-++}^{+-++}(z_1,z_2,z_3,z_4), \nonumber  \\    
                                &= \frac{1}{12} + \frac{1}{12} \left( G_{34} -G_{23} -G_{24} \right) 
+ \frac{1}{12} G_{234} - D(z_1,z_2,z_3,z_4), \nonumber  \\                            
C_2(z_1,z_2,z_3,z_4) &= P_{+-+}^{-++}(z_2,z_3,z_4)-P_{++-+}^{+-++}(z_1,z_2,z_3,z_4),\nonumber  \\
                                &= \frac{1}{12} + \frac{1}{6} G_{12} - E(z_1,z_2,z_3,z_4) + \frac{i}{24} 
\left( G_{123}^A - G_{234}^A \right), \nonumber \\  
C_3(z_1,z_2,z_3,z_4)  & = P_{++-}^{-++}(z_2,z_3,z_4)-P_{+++-}^{+-++}(z_1,z_2,z_3,z_4) \nonumber  \\
                      & = \frac{1}{12} + \frac{1}{6} G_{24} - \frac{1}{6} G_{234} - B(z_4,z_3,z_2,z_1) 
- i \left\{ Y(z_4,z_3,z_2,z_1) - \frac{1}{12} G_{234}^A \right\}, \nonumber  \\
C_4(z_1,z_2,z_3,z_4)  & = C_3(-z_4,-z_3,-z_2,-z_1) \nonumber \\
                      & = \frac{1}{12} + \frac{1}{6} G_{13} - \frac{1}{6} G_{123} - B(z_1,z_2,z_3,z_4) 
+ i \left\{ Y(z_1,z_2,z_3,z_4) - \frac{1}{12} G_{123}^A \right\}, \nonumber  \\
C_5(z_1,z_2,z_3,z_4)  &= \left( 1+ \frac{i}{z_{12}} \right) C_3(z_2,z_1,z_3,z_4) 
- \frac{i}{z_{12}} C_3(z_1,z_2,z_3,z_4), \nonumber  \\ 
C_6(z_1,z_2,z_3,z_4)  &= P_4(z_1,z_2,z_3,z_4) - \sum_{j=1}^{5} C_j(z_1,z_2,z_3,z_4), \nonumber  
\end{align}
\begin{align}
C_7(z_1,z_2,z_3,z_4)  &= P_{-+-}^{-+-}(z_1,z_2,z_3) - P_{-+--}^{-+--}(z_1,z_2,z_3,z_4) \nonumber  \\
                      &= P_{+-+}^{+-+}(z_1,z_2,z_3) - P_{+-++}^{+-++}(z_1,z_2,z_3,z_4) \nonumber  \\
                      &= \frac{1}{12}+\frac{1}{12} \left( G_{13} - G_{23} - G_{12} \right) - \frac{1}{12} G_{123} - D(z_1,z_2,z_3,z_4), \nonumber   \\
C_8(z_1,z_2,z_3,z_4)  &= P_{++-}^{+-+}(z_2,z_3,z_4) - P_{+++-}^{++-+}(z_1,z_2,z_3,z_4) \nonumber  \\
                      &= \frac{1}{12}+\frac{1}{6} G_{34} - A(z_4,z_3,z_2,z_1) - i \left\{ X(z_4,z_3,z_2,z_1)+\frac{1}{12} G_{234}^A \right\},  \nonumber \\
C_9(z_1,z_2,z_3,z_4)  &= C_8(-z_4,-z_3,-z_2,-z_1) \nonumber  \\
                      &= \frac{1}{12}+\frac{1}{6} G_{12} - A(z_1,z_2,z_3,z_4) + i \left\{ X(z_1,z_2,z_3,z_4)+\frac{1}{12} G_{123}^A \right\},  \nonumber \\    
C_{10}(z_1,z_2,z_3,z_4) &= C_6(z_1,z_3,z_2,z_4) + \frac{i}{z_{32}} \left\{ C_5(z_1,z_3,z_2,z_4)-C_5(z_1,z_2,z_3,z_4) \right\}, \nonumber  \\
C_{11}(z_1,z_2,z_3,z_4) &= P_{-++}^{-++}(z_1,z_2,z_3) - P_{-+++}^{-+++}(z_1,z_2,z_3,z_4) \nonumber  \\
                        &= -\frac{1}{6} - \frac{1}{12} (G_{12}+G_{13}+G_{24}+G_{34}) + \frac{1}{12}(G_{123}+G_{234}) +P_4(z_1,z_2,z_3,z_4), \nonumber  \\
C_{12}(z_1,z_2,z_3,z_4) &=  C_{10}(z_1,z_2,z_4,z_3) + \frac{i}{z_{43}} \left\{ C_9(z_1,z_2,z_4,z_3)- C_9(z_1,z_2,z_3,z_4) \right\}.          
\end{align}

Among the correlation functions above, let us pick up a diagonal one ${P_{--++}^{--++}(z_1,z_2,z_3,z_4)}$, since it has a relatively simple explicit form.
\begin{align}
P_{--++}^{--++}(z_1,z_2,z_3,z_4)& 
= \frac{1}{30} + A_{4,1}^{(2)}(z_1,z_2,z_3,z_4) G(z_{12}) + 
\left\{ A_{4,1}^{(2)}(z_1,z_3,z_2,z_4) - \frac{1}{12} \right\} G(z_{13}) \nonumber  \\ 
& + \left\{ A_{4,1}^{(2)}(z_1,z_4,z_3,z_2) - \frac{1}{12} \right\} G(z_{14}) 
+ \left\{ A_{4,1}^{(2)}(z_3,z_2,z_1,z_4) - \frac{1}{12} \right\} G(z_{23})  \nonumber  \\
& + \left\{ A_{4,1}^{(2)}(z_4,z_2,z_3,z_1) - \frac{1}{12} \right\} G(z_{24}) 
+ A_{4,1}^{(2)}(z_4,z_3,z_2,z_1) G(z_{34}) \nonumber  \\
& + A_{4,2}^{(2)}(z_1,z_2,z_3,z_4) G(z_{12}) G(z_{34}) 
+ A_{4,2}^{(2)}(z_1,z_3,z_2,z_4) G(z_{13}) G(z_{24}) \nonumber  \\
& + A_{4,2}^{(2)}(z_1,z_4,z_3,z_2) G(z_{14}) G(z_{23}), \nonumber 
\end{align}
with
\begin{align}
A_{4,i}^{(2)}(z_1,z_2,z_3,z_4) &= \frac{Q_{4,i}^{(2)}(z_1,z_2,z_3,z_4)}{z_{13}z_{14} z_{23} z_{24}}, \ \ \ \ (i=1,2), \nonumber  \\
Q_{4,2}^{(2)}(z_1,z_2,z_3,z_4) &= \frac{1}{36} \left\{(z_{13} z_{24}+1)(z_{14} z_{23}+1) + \frac{2}{5} 
\left( z_{12}^2+\frac{3}{2} \right) \left( z_{34}^2 +\frac{3}{2} \right) -\frac{5}{2} \right\}, \nonumber  \\
Q_{4,1}^{(2)}(z_1,z_2,z_3,z_4) &= 2 Q_{4,2}^{(2)}(z_1,z_2,z_3,z_4) - \frac{1}{60}(z_{12}^2-1)(z_{34}^2+1).
\label{corrN4sec2}
\end{align}
The correlation functions in the remaining sector ${\e_1+\e_2+\e_3+\e_4 = \e_1'+\e_2'+\e_3'+\e_4'=-2}$ are obtained from 
those in ${\e_1+\e_2+\e_3+\e_4 = \e_1'+\e_2'+\e_3'+\e_4'=2}$ by the negation relation 
\begin{equation}
P_{\e_1 \e_2 \e_3 \e_4}^{\e'_1 \e'_2 \e'_3 \e'_4}(z_1,z_2,z_3,z_4) 
= P_{-\e_1,-\e_2,-\e_3,-\e_4}^{-\e'_1,-\e'_2,-\e'_3,-\e'_4}(z_1,z_2,z_3,z_4).
\end{equation}
In this way we can get all the inhomogeneous correlation functions for ${n=4}$. 

Now we shortly consider the diagonal elements of the correlation functions 
${P_{\e_1 \e_2 \e_3 \e_4}^{\e_1 \e_2 \e_3 \e_4}(z_1,z_2,z_3,z_4)}$. We claim they can be 
expressed only in terms of the EFP's, ${P_2(z_1,z_2), P_3(z_1,z_2,z_3), P_4(z_1,z_2,z_3,z_4)}$ and 
${P_{--++}^{--++}(z_1,z_2,z_3,z_4)}$. For example, we have 
\begin{align}
P_{-+++}^{-+++}(z_1,z_2,z_3,z_4) &= P_3(z_2,z_3,z_4) -P_4(z_1,z_2,z_3,z_4), \nonumber  \\
P_{+-++}^{+-++}(z_1,z_2,z_3,z_4) &= P_2(z_3,z_4) -P_3(z_2,z_3,z_4)-P_{--++}^{--++}(z_1,z_2,z_3,z_4), \nonumber  \\
P_{+-+-}^{+-+-}(z_1,z_2,z_3,z_4) &= \frac{1}{2}- P_2(z_1,z_2)
-P_2(z_2,z_3)-P_2(z_3,z_4) \nonumber \\
& +P_3(z_1,z_2,z_3)+P_3(z_2,z_3,z_4)+P_{--++}^{--++}(z_1,z_2,z_3,z_4), \nonumber  \\
P_{+--+}^{+--+}(z_1,z_2,z_3,z_4) &= P_2(z_2,z_3)-P_3(z_1,z_2,z_3)-P_3(z_2,z_3,z_4) +P_4(z_1,z_2,z_3,z_4). 
\end{align}
Then, especially, the inhomogeneous 
generalization of the third-neighbor correlator ${\left\langle S_1^{z} S_4^{z} \right\rangle}$
is represented  
\begin{align}
& \left\langle S_1^{z} S_4^{z} \right\rangle (z_1,z_2,z_3,z_4) \nonumber  \\
&= \frac{1}{4} \Big\{ P_{++++}^{++++}(z_1,z_2,z_3,z_4)-P_{-+++}^{-+++}(z_1,z_2,z_3,z_4) 
+P_{+-++}^{+-++}(z_1,z_2,z_3,z_4) - P_{--++}^{--++}(z_1,z_2,z_3,z_4) \nonumber  \\
& \ \ \ \ \ \ +P_{++-+}^{++-+}(z_1,z_2,z_3,z_4) - P_{-+-+}^{-+-+}(z_1,z_2,z_3,z_4) 
+P_{+--+}^{+--+}(z_1,z_2,z_3,z_4) - P_{---+}^{---+}(z_1,z_2,z_3,z_4) \nonumber  \\
& \ \ \ \ \ \ -P_{+++-}^{+++-}(z_1,z_2,z_3,z_4) + P_{-++-}^{-++-}(z_1,z_2,z_3,z_4) 
-P_{+-+-}^{+-+-}(z_1,z_2,z_3,z_4) + P_{--+-}^{--+-}(z_1,z_2,z_3,z_4) \nonumber  \\
& \ \ \ \ \ \ -P_{++--}^{++--}(z_1,z_2,z_3,z_4) + P_{-+--}^{-+--}(z_1,z_2,z_3,z_4) 
-P_{+---}^{+---}(z_1,z_2,z_3,z_4) + P_{----}^{----}(z_1,z_2,z_3,z_4) \Big\} \nonumber  \\
& = -\frac{1}{4} + P_2(z_1,z_2)+P_2(z_2,z_3)+P_2(z_3,z_4)- 2 \left\{ P_3(z_1,z_2,z_3)+P_3(z_2,z_3,z_4) \right\} \nonumber \\
& \ \ \ \ \ \ + 2 \left\{ P_4(z_1,z_2,z_3,z_4)-P_{--++}^{--++}(z_1,z_2,z_3,z_4) \right\} \nonumber  \\
& = \frac{1}{12} + \frac{1}{6} G_{14} + \frac{1}{6} \left( G_{123}+G_{234} \right) 
+ A_{4,1}^{(3)}(z_1,z_2,z_3,z_4) G_{12} + A_{4,1}^{(3)}(z_1,z_3,z_2,z_4) G_{13}   \nonumber  \\ 
& \ \ \ \ \ \ + A_{4,1}^{(3)}(z_1,z_4,z_3,z_2) G_{14} + A_{4,1}^{(3)}(z_3,z_2,z_1,z_4) G_{23} + A_{4,1}^{(3)}(z_4,z_2,z_3,z_1) G_{24} \nonumber  \\
& \ \ \ \ \ \ + A_{4,1}^{(3)}(z_4,z_3,z_2,z_1) G_{34} + A_{4,2}^{(3)}(z_1,z_2,z_3,z_4) G_{12} G_{34} 
+ A_{4,2}^{(3)}(z_1,z_3,z_2,z_4) G_{13} G_{24} \nonumber  \\
& \ \ \ \ \ \ + A_{4,2}^{(3)}(z_1,z_4,z_3,z_2) G_{14} G_{23}, \nonumber  
\end{align} 
where
\begin{align}
A_{4,2}^{(3)}(z_1,z_2,z_3,z_4) &= 2 \left\{ A_{4,2}^{(0)}(z_1,z_2,z_3,z_4)  -A_{4,2}^{(3)}(z_1,z_2,z_3,z_4)  \right\} \nonumber  \\
                               &= \frac{1}{9} \left( \frac{2}{z_{13} z_{14} z_{23} z_{24}} -\frac{1}{z_{14} z_{23}} -\frac{1}{z_{13} z_{24}} \right), 
                               \nonumber  \\
A_{4,1}^{(3)}(z_1,z_2,z_3,z_4) &= 2 \left\{ A_{4,1}^{(0)}(z_1,z_2,z_3,z_4)  -A_{4,1}^{(3)}(z_1,z_2,z_3,z_4)  \right\} \nonumber  \\
                               &= \frac{1}{18} \left( \frac{5}{z_{13} z_{14} z_{23} z_{24}} -\frac{1}{z_{14} z_{23}} -\frac{1}{z_{13} z_{24}} 
- \frac{3}{z_{14} z_{24}} - \frac{3}{z_{13} z_{23}} \right).                               
\end{align}
By taking the homogeneous limit ${z_i \to 0}$, we recover the result (\ref{corr_n4}).
\begin{align}
& \lim_{z_i \to 0} \left\langle S_1^{z} S_4^{z} \right\rangle (z_1,z_2,z_3,z_4) 
= \frac{1}{12} - \frac{1}{3} \zeta_a(1) + 2\left\{ \zeta_a(1)-\zeta_a(3) \right\} \nonumber  \\
& + \Big\{ - \frac{14}{3} \zeta_a(1) + \frac{92}{9} \zeta_a(3) - \frac{50}{9} \zeta_a(5) 
- \frac{56}{9} \zeta_a(1) \zeta_a(3) - \frac{8}{3} \zeta_a(3)^2 + \frac{80}{9} \zeta_a(1) \zeta_a(5) \Big\} \nonumber  \\
&= \frac{1}{12} - 3 \zeta_a(1) + \frac{74}{9} \zeta_a(3) 
- \frac{50}{9} \zeta_a(5) - \frac{56}{9} \zeta_a(1) \zeta_a(3)- \frac{8}{3} \zeta_a(3)^2 + \frac{80}{9} \zeta_a(1) \zeta_a(5) \nonumber  \\
&= \frac{1}{12} - 3 \ln 2 + \frac{37}{6} \zeta(3) - \frac{125}{24} \zeta(5) 
- \frac{14}{3} \ln 2 \cdot \zeta(3) - \frac{3}{2} \zeta(3)^2 + \frac{25}{3} \ln 2 \cdot \zeta(5). 
\end{align}
Other correlation functions in (\ref{corr_n4}) are reproduced in a similar way.

\section{Correlation functions in the case $n=5$}

Following the scheme in the previous sections we could further obtain all the correlation 
functions for the ${XXX}$ chain on five lattice sites. In fact we have calculated two inhomogeneous 
correlation functions ${P_{++++-}^{-++++}(z_1,z_2,z_3,z_4,z_5)}$ and ${P_{+++--}^{--+++}(z_1,z_2,z_3,z_4,z_5)}$ 
from the first principle relations. Other correlation functions are derived by the intertwining relations, etc...,
as before.  The calculations have been performed by the heavy use of {\it Mathematica} and the obtained results are 
too complicated to be described thoroughly in this paper. Therefore here we shall omit to describe them.  We, however, give explicit form of 
the inhomogeneous generalization of the fourth-neighbor correlation function ${\langle S_{1}^z S_{5}^z \rangle (z_1,z_2,z_3,z_4,z_5)}$. 
\begin{align}
& \langle S_{1}^z S_{5}^z \rangle (z_1,z_2,z_3,z_4,z_5) \equiv \frac{1}{4} \sum_{\e_j=\pm} \e_1 \e_5 
P_{\e_1 \e_2 \e_3 \e_4 \e_5}^{\e_1 \e_2 \e_3 \e_4 \e_5} (z_1,z_2,z_3,z_4,z_5) \nonumber  \\
& = \frac{1}{12} + A_{5,1}(z_1,z_2,z_3,z_4,z_5) G_{12} +  A_{5,1}(z_1,z_3,z_2,z_4,z_5) G_{13} 
  + A_{5,1}(z_1,z_4,z_3,z_2,z_5) G_{14} \nonumber  \\
& +  A_{5,1}(z_5,z_2,z_3,z_4,z_1) G_{25} +  A_{5,1}(z_5,z_3,z_2,z_4,z_1) G_{35} 
  +  A_{5,1}(z_5,z_4,z_3,z_2,z_1) G_{45}  \nonumber  \\
& + A_{5,2}(z_2,z_3,z_1,z_4,z_5) G_{23}  + A_{5,2}(z_2,z_4,z_1,z_3,z_5) G_{24}  
+ A_{5,2}(z_3,z_4,z_1,z_2,z_5) G_{34} \nonumber  \\
& + A_{5,3}(z_1,z_5,z_2,z_3,z_4) G_{15} + A_{5,4}(z_1,z_2,z_3,z_4,z_5) G_{12} G_{34} 
+ A_{5,4}(z_1,z_3,z_2,z_4,z_5) G_{13} G_{24} \nonumber  \\
& + A_{5,4}(z_1,z_4,z_2,z_3,z_5) G_{14} G_{23} + A_{5,4}(z_5,z_2,z_3,z_4,z_1) G_{25} G_{34} 
+ A_{5,4}(z_5,z_3,z_2,z_4,z_1) G_{35} G_{24} \nonumber  \\
& + A_{5,4}(z_5,z_4,z_2,z_3,z_1) G_{45} G_{23} + A_{5,5}(z_1,z_2,z_3,z_5,z_4) G_{12} G_{35} 
+ A_{5,5}(z_1,z_3,z_2,z_5,z_4) G_{13} G_{25} \nonumber  \\
& + A_{5,4}(z_1,z_2,z_4,z_5,z_3) G_{12} G_{45} + A_{5,5}(z_1,z_4,z_2,z_5,z_3) G_{14} G_{25} 
+ A_{5,5}(z_1,z_3,z_4,z_5,z_2) G_{13} G_{45} \nonumber  \\
& + A_{5,4}(z_1,z_4,z_3,z_5,z_2) G_{14} G_{35} + A_{5,5}(z_1,z_5,z_2,z_3,z_4) G_{15} G_{23} 
+ A_{5,5}(z_1,z_5,z_3,z_4,z_2) G_{15} G_{34} \nonumber  \\
& + A_{5,5}(z_1,z_5,z_2,z_4,z_3) G_{15} G_{24},  \nonumber  
\end{align} 
where
\begin{align}
A_{5,i}(z_1,z_2,z_3,z_4,z_5) = \begin{cases}
\displaystyle{\frac{Q_{5,i}(z_1,z_2,z_3,z_4,z_5)}{z_{13} z_{14} z_{15} z_{23} z_{24} z_{25}}}, & (i=1,2,3) \nonumber  \\ \\
\displaystyle{\frac{Q_{5,i}(z_1,z_2,z_3,z_4,z_5)}{z_{13} z_{14} z_{15} z_{23} z_{24} z_{25} z_{34} z_{35}}}, & (i=4,5) 
\end{cases}
\end{align}
with 
\begin{align}
Q_{5,1}(z_1,z_2,z_3,z_4,z_5) = & \frac{1}{36} \big[ (z_{15}+z_{25})(z_{13}+z_{24})(2 - z_{13} z_{14} - z_{14} z_{23})  - 6 (z_{13} z_{14}-2)(z_{23} z_{24}-2) \nonumber \\
& +z_{12}^2 (z_{13} z_{23} + z_{14} z_{24} - 16)+ 10 z_{34}^2 -2 \big], \nonumber  \\
Q_{5,2}(z_1,z_2,z_3,z_4,z_5) = & Q_{5,1}(z_1,z_2,z_3,z_4,z_5) 
                               + \frac{1}{18} z_{13} z_{23} \big[ 5 z_{14} z_{24} + z_{45}(z_{14}+z_{24}) -5 \big], \nonumber  \\
Q_{5,3}(z_1,z_2,z_3,z_4,z_5) = & Q_{5,1}(z_1,z_2,z_3,z_4,z_5) 
                               + \frac{1}{18} z_{15} z_{25} \big[ 3 z_{13} z_{14} z_{23} z_{24} -4(z_{13} z_{24}+z_{14} z_{23}) - 3 z_{34}^2 +5 \big], \nonumber  \\
Q_{5,4}(z_1,z_2,z_3,z_4,z_5) = & \frac{1}{18} \big[ 2 (z_{13} z_{24} + z_{14} z_{23})z_{15} z_{25} - z_{13} z_{23} z_{24} z_{45} - z_{14} z_{23} z_{24} z_{35}
- z_{13} z_{14} z_{23} z_{45} \nonumber  \\
& - z_{13} z_{14} z_{24} z_{35} + z_{12}^2 z_{34}^2 + 6 (z_{12}^2 +z_{34}^2) - \frac{3}{2} \left( z_{13}^2 + z_{14}^2 + z_{23}^2 + z_{24}^2 \right) \nonumber  \\
& - \left( z_{15}^2 +z_{25}^2 + z_{35}^2 + z_{45}^2 \right) + 12  \big], \nonumber  \\
Q_{5,5}(z_1,z_2,z_3,z_4,z_5) = & Q_{5,4}(z_1,z_2,z_3,z_4,z_5) + \frac{1}{9} ( 2 - z_{13} z_{24} - z_{14} z_{23}) z_{15} z_{25} z_{35} z_{45}.                             
\end{align}
By taking the homogeneous limit, we obtain the fourth-neighbor correlation function for the homogeneous Heisenberg model as follows. 
\begin{align}
\left\langle S_j^{z} S_{j+4}^{z} \right\rangle & = \lim_{z_{i} \to 0} \left\langle S_1^{z} S_{5}^{z} \right\rangle (z_1,z_2,z_3,z_4,z_5) \nonumber  \\
& = \frac{1}{12} - \frac{16}{3} \zeta_a(1) + \frac{290}{9} \zeta_a(3) - 72 \zeta_a(1) \cdot \zeta_a(3) 
- \frac{1172}{9} \zeta_a(3)^2 - \frac{700}{9} \zeta_a(5) \nonumber \\ 
&  + \frac{4640}{9} \zeta_a(1) \cdot \zeta_a(5)  - \frac{220}{9} \zeta_a(3) \cdot \zeta_a(5) - \frac{400}{3} \zeta_a(5)^2 + \frac{455}{9} \zeta_a(7)  \nonumber  \\
& - \frac{3920}{9} \zeta_a(1) \cdot \zeta_a(7) + 280  \zeta_a(3) \cdot \zeta_a(7)  \nonumber  \\ 
& = \frac{1}{12} - \frac{16}{3}
\ln 2  + \frac{145}{6} \zeta(3) - 54 \ln 2 \cdot \zeta(3) 
- \frac{293}{4} \zeta(3)^2 - \frac{875}{12} \zeta(5) \nonumber \\ 
&  + \frac{1450}{3} \ln 2 \cdot \zeta(5)  - \frac{275}{16} \zeta(3) \cdot \zeta(5) - \frac{1875}{16} \zeta(5)^2 + \frac{3185}{64} \zeta(7)  \nonumber  \\
& - \frac{1715}{4} \ln 2 \cdot \zeta(7) + \frac{6615}{32}  \zeta(3) \cdot \zeta(7) 
= 0.034652776982 \cdots. \label{fourth-neighbor}
\end{align}
This is one of the main new results of this paper. Furthermore we have obtained all the other correlation functions for ${n=5}$. 
We present below the homogeneous limit of independent ones. 
\begin{align}
\left\langle S_j^{x} S_{j+1}^{x} S_{j+2}^{z} S_{j+4}^{z} \right\rangle 
& = \frac{1}{240} + \frac{1}{12} \ln 2  - \frac{517}{480} \zeta(3) + \frac{25}{12} \ln 2 \cdot \zeta(3) 
+ \frac{203}{80} \zeta(3)^2 + \frac{1525}{384} \zeta(5) \nonumber  \\ 
& - \frac{215}{12} \ln 2 \cdot \zeta(5) + \frac{5}{16} \zeta(3) \cdot \zeta(5) + \frac{125}{32} \zeta(5)^2 - \frac{735}{256} \zeta(7) \nonumber  \\
& + \frac{1029}{64} \ln 2 \cdot \zeta(7) -  \frac{441}{64}  \zeta(3) \cdot \zeta(7) =-0.0098924350847 \cdots, 
\end{align}
\begin{align}
\left\langle S_j^{x} S_{j+1}^{z} S_{j+2}^{x} S_{j+4}^{z} \right\rangle 
& = \frac{1}{240} - \frac{1}{4} \ln 2  + \frac{301}{160} \zeta(3) - \frac{9}{2} \ln 2 \cdot \zeta(3) 
- \frac{1079}{160} \zeta(3)^2 - \frac{975}{128} \zeta(5) \nonumber  \\
& + \frac{365}{8} \ln 2 \cdot \zeta(5) - \frac{185}{128} \zeta(3) \cdot \zeta(5) - \frac{1375}{128} \zeta(5)^2 + \frac{735}{128} \zeta(7) \nonumber  \\
& - \frac{1323}{32} \ln 2 \cdot \zeta(7) +  \frac{4851}{256}  \zeta(3) \cdot \zeta(7)= 0.0027889733995 \cdots,  
\end{align}
\begin{align}
\left\langle S_j^{x} S_{j+1}^{z} S_{j+2}^{z} S_{j+4}^{x} \right\rangle 
& = \frac{1}{240} - \frac{5}{12} \ln 2  + \frac{569}{240} \zeta(3) - \frac{61}{12} \ln 2 \cdot \zeta(3) 
- \frac{1109}{160} \zeta(3)^2 - \frac{775}{96} \zeta(5) \nonumber \\ 
& + \frac{140}{3} \ln 2 \cdot \zeta(5)  - \frac{185}{128} \zeta(3) \cdot \zeta(5) 
- \frac{1375}{128} \zeta(5)^2 + \frac{735}{128} \zeta(7) \nonumber  \\ 
&- \frac{1323}{32} \ln 2 \cdot \zeta(7) +  \frac{4851}{256}  \zeta(3) \cdot \zeta(7)=-0.00505666089481 \cdots,   
\end{align}
\begin{align}
\left\langle S_j^{x} S_{j+1}^{x} S_{j+3}^{z} S_{j+4}^{z} \right\rangle 
& = \frac{1}{240} + \frac{1}{4} \ln 2  - \frac{419}{160} \zeta(3) + \frac{35}{6} \ln 2 \cdot \zeta(3) 
+ \frac{663}{80} \zeta(3)^2 + \frac{4445}{384} \zeta(5) \nonumber \\ 
& - \frac{115}{2} \ln 2 \cdot \zeta(5) + \frac{85}{64} \zeta(3) \cdot \zeta(5) + \frac{1625}{128} \zeta(5)^2 - \frac{2303}{256} \zeta(7) \nonumber  \\
&+ \frac{833}{16} \ln 2 \cdot \zeta(7) 
 -  \frac{5733}{256}  \zeta(3) \cdot \zeta(7) = 0.01857662093837 \cdots, 
\end{align}
\begin{align} 
\left\langle S_j^{x} S_{j+1}^{z} S_{j+3}^{x} S_{j+4}^{z} \right\rangle & = \frac{1}{240} - \frac{5}{12} \ln 2  + \frac{1883}{480} \zeta(3) - \frac{28}{3} \ln 2 \cdot \zeta(3) 
- \frac{551}{40} \zeta(3)^2 - \frac{2285}{128} \zeta(5) \nonumber  \\
& + \frac{1135}{12} \ln 2 \cdot \zeta(5)  - \frac{165}{64} \zeta(3) \cdot \zeta(5) - \frac{1375}{64} \zeta(5)^2 + \frac{3577}{256} \zeta(7) \nonumber  \\
&- \frac{343}{4} \ln 2 \cdot \zeta(7) +  \frac{4851}{128}  \zeta(3) \cdot \zeta(7) =-0.00108936897291 \cdots,   
\end{align}
\begin{align}
\left\langle S_j^{x} S_{j+1}^{z} S_{j+3}^{z} S_{j+4}^{x} \right\rangle & = \frac{1}{240} - \frac{1}{2} \ln 2  + \frac{497}{120} \zeta(3) - \frac{28}{3} \ln 2 \cdot \zeta(3) 
- \frac{551}{40} \zeta(3)^2 - \frac{1155}{64} \zeta(5) \nonumber  \\
&+ \frac{1135}{12} \ln 2 \cdot \zeta(5) - \frac{165}{64} \zeta(3) \cdot \zeta(5) - \frac{1375}{64} \zeta(5)^2 + \frac{3577}{256} \zeta(7) \nonumber  \\
&- \frac{343}{4} \ln 2 \cdot \zeta(7) +  \frac{4851}{128}  \zeta(3) \cdot \zeta(7)=0.001573361370145 \cdots.    
\end{align}
As a confirmation of our results, we have evaluated the numerical values of the correlation functions for the finite system size ${N}$ up to 32 (Table 4). 
We have further applied an extrapolation to this data such as ${c_\infty + c_1/N^2 + c_2/N^4 +  c_3/N^6}$, and obtained the estimated values ${c_\infty}$ in the limit ${N \to \infty}$. 
One can clearly observe our analytical results coincide with the extrapolated values with extremely high accuracy (more than 4 digits). 


\begin{table}[htbp]
\begin{center}
\caption{\small{Numerical values of correlation functions for finite systems}}
\label{table1}
\begin{tabular}{@{\hspace{\tabcolsep}\extracolsep{\fill}}crrrrr} \hline
           & $N$=26 \ \ \  & $N$=28 \ \ \  & $N$=30 \ \ \  & $N$=32 \ \ \  & ${c_\infty}$ (${N=\infty}$)   \\ \hline
${\left\langle S_j^{z}S_{j+4}^{z} \right\rangle}$               & 0.0357233 & 0.0355713 & 0.0354497 & 0.0353508 & 0.0346535 \\
${\left\langle S_j^{x}S_{j+1}^{x}S_{j+2}^{z}S_{j+4}^{z} \right\rangle}$ &-0.0099654 &-0.0099553 &-0.0099471 &-0.0099405 &-0.0098925 \\
${\left\langle S_j^{x}S_{j+1}^{z}S_{j+2}^{x}S_{j+4}^{z} \right\rangle}$ & 0.0028572 & 0.0028477 & 0.0028400 & 0.0028337 & 0.0027890 \\
${\left\langle S_j^{x}S_{j+1}^{z}S_{j+2}^{z}S_{j+4}^{x} \right\rangle}$ &-0.0052334 &-0.0052083 &-0.0051882 &-0.0051719 &-0.0050564 \\ 
${\left\langle S_j^{x}S_{j+1}^{x}S_{j+3}^{z}S_{j+4}^{z} \right\rangle}$ & 0.0185725 & 0.0185731 & 0.0185737 & 0.0185741 & 0.0185766 \\ 
${\left\langle S_j^{x}S_{j+1}^{z}S_{j+3}^{x}S_{j+4}^{z} \right\rangle}$ &-0.0011377 &-0.0011309 &-0.0011254 &-0.0011209 &-0.0010893 \\
${\left\langle S_j^{x}S_{j+1}^{z}S_{j+3}^{z}S_{j+4}^{x} \right\rangle}$ & 0.0016526 & 0.0016413 & 0.0016323 & 0.0016250 & 0.0015732 \\ 
\hline
\end{tabular}
\end{center}
\end{table}
We make several comments on our results. First, the two other diagonal correlation functions are given as a sum of non-diagonal ones 
similarly in the case of four lattice sites, 
\begin{align}
\left\langle S_j^{z} S_{j+1}^{z} S_{j+2}^{z} S_{j+4}^{z} \right\rangle & = \left\langle S_j^{x} S_{j+1}^{x} S_{j+2}^{z} S_{j+4}^{z} \right\rangle 
+\left\langle S_j^{x} S_{j+1}^{z} S_{j+2}^{x} S_{j+4}^{z} \right\rangle +\left\langle S_j^{x} S_{j+1}^{z} S_{j+2}^{z} S_{j+4}^{x} \right\rangle 
\nonumber \\
& = \frac{1}{80} - \frac{7}{12} \ln 2  + \frac{127}{40} \zeta(3) - \frac{15}{2} \ln 2 \cdot \zeta(3) 
- \frac{891}{80} \zeta(3)^2 - \frac{375}{32} \zeta(5)   \nonumber \\ 
& + \frac{595}{8} \ln 2 \cdot \zeta(5)- \frac{165}{64} \zeta(3) \cdot \zeta(5) - \frac{1125}{64} \zeta(5)^2 + \frac{2205}{256} \zeta(7) \nonumber  \\
& - \frac{4263}{64} \ln 2 \cdot \zeta(7) 
 +  \frac{3969}{128}  \zeta(3) \cdot \zeta(7) =-0.0121601225799 \cdots,   
\end{align} 
\begin{align}
\left\langle S_j^{z} S_{j+1}^{z} S_{j+3}^{z} S_{j+4}^{z} \right\rangle & = \left\langle S_j^{x} S_{j+1}^{x} S_{j+3}^{z} S_{j+4}^{z} \right\rangle 
+\left\langle S_j^{x} S_{j+1}^{z} S_{j+3}^{x} S_{j+4}^{z} \right\rangle +\left\langle S_j^{x} S_{j+1}^{z} S_{j+3}^{z} S_{j+4}^{x} \right\rangle 
\nonumber \\
& = \frac{1}{80} - \frac{2}{3} \ln 2  + \frac{1307}{240} \zeta(3) - \frac{77}{6} \ln 2 \cdot \zeta(3) 
- \frac{1541}{80} \zeta(3)^2 - \frac{2335}{96} \zeta(5)  \nonumber \\ 
& + \frac{395}{3} \ln 2 \cdot \zeta(5) - \frac{245}{64} \zeta(3) \cdot \zeta(5) - \frac{3875}{128} \zeta(5)^2 + \frac{4851}{256} \zeta(7) \nonumber  \\
& - \frac{1911}{16} \ln 2 \cdot \zeta(7) 
 +  \frac{13671}{256}  \zeta(3) \cdot \zeta(7) =0.0190606133356 \cdots.  
\end{align}

The expression for ${P(5)}$ (\ref{P5}) is reproduced as
\begin{align}
P(5) &= \frac{1}{32} + \frac{1}{2} \left\langle S_j^{z} S_{j+1}^{z} \right\rangle + \frac{3}{8} \left\langle S_j^{z} S_{j+2}^{z} \right\rangle 
+ \frac{1}{4} \left\langle S_j^{z} S_{j+3}^{z} \right\rangle + \frac{1}{8} \left\langle S_j^{z} S_{j+4}^{z} \right\rangle \nonumber  \\
& + \left\langle S_j^{z} S_{j+1}^{z} S_{j+2}^{z} S_{j+3}^{z} \right\rangle 
+ \left\langle S_j^{z} S_{j+1}^{z} S_{j+3}^{z} S_{j+4}^{z} \right\rangle +
\frac{1}{2} \left\langle S_j^{z} S_{j+1}^{z} S_{j+3}^{z} S_{j+4}^{z} \right\rangle \nonumber \\
& = \frac{1}{6} - \frac{10}{3} \ln 2  + \frac{281}{24} \zeta(3) - \frac{45}{2} \ln 2 \cdot \zeta(3) 
- \frac{489}{16} \zeta(3)^2 - \frac{6775}{192} \zeta(5)  \nonumber \\ 
& + \frac{1225}{6} \ln 2 \cdot \zeta(5) - \frac{425}{64} \zeta(3) \cdot \zeta(5) - \frac{12125}{256} \zeta(5)^2 + \frac{6223}{256} \zeta(7) \nonumber  \\
& - \frac{11515}{64} \ln 2 \cdot \zeta(7) 
 +  \frac{42777}{512}  \zeta(3) \cdot \zeta(7) = 2.0117259 \cdots \times 10^{-6}. \nonumber 
\end{align}
In fact, arbitrary correlation functions on five lattice sites are similarly calculated.

Finally we notice that the ``next-nearest chiral correlator"  simplifies drastically as 
\begin{align}
\left\langle \left({\bf S}_j \times {\bf S}_{j+1} \right) \cdot  \left({\bf S}_{j+3} \times {\bf S}_{j+4} \right) \right\rangle  
&=6 \left( \left\langle S_j^{x} S_{j+1}^{z} S_{j+3}^{x} S_{j+4}^{z} \right\rangle -\left\langle S_j^{x} S_{j+1}^{z} S_{j+3}^{z} S_{j+4}^{x} \right\rangle \right) 
\nonumber  \\
&= \frac{1}{2} \ln 2 - \frac{21}{16} \zeta(3) + \frac{75}{64} \zeta(5)  \nonumber \\
&= -0.015976382058 \cdots, 
\end{align}
as well as a similar correlator 
\begin{align}
\left\langle \left({\bf S}_j \times {\bf S}_{j+1} \right) \cdot  \left({\bf S}_{j+2} \times {\bf S}_{j+4} \right) \right\rangle  
&=6 \left( \left\langle S_j^{x} S_{j+1}^{z} S_{j+2}^{x} S_{j+4}^{z} \right\rangle -\left\langle S_j^{x} S_{j+1}^{z} S_{j+2}^{z} S_{j+4}^{x} \right\rangle \right) 
\nonumber \\
&= \ln 2 - \frac{47}{16} \zeta(3) + \frac{7}{2} \ln 2 \cdot \zeta(3) + \frac{9}{8} \zeta(3)^2 + \frac{175}{64} \zeta(5) - \frac{25}{4} \ln 2 \cdot \zeta(5) \nonumber \\
&= 0.04707380576617 \cdots.  
\end{align}
Recall that a similar simplification occurs in the case of ``nearest chiral correlator"  \cite{MT}
\begin{align}
\left\langle \left({\bf S}_j \times {\bf S}_{j+1} \right) \cdot  \left({\bf S}_{j+2} \times {\bf S}_{j+3} \right) \right\rangle  
&=6 \left( \left\langle S_j^{x} S_{j+1}^{z} S_{j+2}^{x} S_{j+3}^{z} \right\rangle -\left\langle S_j^{x} S_{j+1}^{z} S_{j+2}^{z} S_{j+3}^{x} \right\rangle \right) 
\nonumber  \\
&= \frac{1}{2} \ln 2 - \frac{3}{8} \zeta(3).  
\end{align}

\section{Summary}
In this paper, we 
calculate correlation functions of the Heisenberg chain 
without magnetic field in the anti-ferromagnetic ground state using
several fundamental functional relations followed from the quantum Knizhnik-Zamolodchikov 
equations. This is the generalization of the method proposed for the emptiness formation probability. 
In fact we have calculated all the correlation functions on five lattice sites. Especially we could get 
the analytic formula for the fourth-neighbor correlation ${\left\langle S_j^{z} S_{j+4}^{z} \right\rangle}$. 
In principle we can continue to calculate ${\left\langle S_j^{z} S_{j+k}^{z} \right\rangle_{k \ge 5}}$, 
which will be reported in the further publications. 

\section{Acknowledgment}
Authors are grateful to F. G{\"o}hmann, M. Jimbo, A. Kl{\"u}mper, V. Korepin, S. Lukyanov, 
T. Miwa, K. Sakai, J. Sato,  F. Smirnov, Y. Takeyama and Z. Tsuboi for useful discussions. 
HB would like to thank  to ISSP of Tokyo University where work on this paper was partially 
done. This work is in part supported by Grant-in-Aid for the Scientific Research (B) no 14340099. 
MS is also supported by Grant-in-Aid for Young Scientists (B) no 14740228.  
 HB was supported by INTAS grant \#00-00561 and by the RFFI grant \#04-01-00352.


\begin{thebibliography}{XXXXXXX}

\bibitem{Heis} W. Heisenberg, Z. Phys. {\bf 49} (9-10)  (1928) 619.

\bibitem{takbook} M.~Takahashi, {\it Thermodynamics of One-Dimensional Solvable 
Models}, Cambridge University Press, Cambridge, 1999.  

\bibitem{B} H.~Bethe, Z. Phys. {\bf 76} (1931) 205. 

\bibitem{H} L.~Hulth\'{e}n, Ark. Mat. Astron. Fysik {\bf A 26} (1939) 1.

\bibitem{ABA} L.~Faddeev,
{\it ``How algebraic Bethe Ansatz works for integrable models"}, 
Les HouchesSession LXIV (1995), Quantum symmetries,
eds A.~Connes, K.~Gawedzki and J.~Zinn-Justin,
Elsevier Science (1998) pp. 149 - 219.

\bibitem{FT} L.~Faddeev and L.~Takhtajan
{\it ``Spectrum and scattering of excitations in the one-dimensional
isotropic Heisenberg model"},
J. Sov. Math. {\bf 24} (1984) 241-267.

\bibitem{bax} R.~Baxter,
{\it Exactly Solved Models in Statistical Mechanics},
Academic Press, New York, 1982.

\bibitem{tak} M.~Takahashi, J. Phys. C: Solid State Phys. 
{\bf 10} (1977) 1289. 

\bibitem{DI} J.~Dittrich and V.~I. Inozemtsev, J. Phys. A {\bf 30} (1997) L623.  

\bibitem{KZ} V.G. Knizhnik and A.B. Zamolodchikov, 
Nucl. Phys. B {\bf 247} (1984) 83. 

\bibitem{book} F.A.~Smirnov,
{\it Form Factors in Completely Integrable Models of Quantum Field Theory.}
Adv. Series in Math. Phys. 14, World Scientific, Singapore, 1992.

\bibitem{sm} F.A. Smirnov, Int. J. Math. Phys. A7 (Suppl. 1B) (1992) 813.

\bibitem{FR} I.~Frenkel, N. Reshetikhin, Commun. Math. Phys. {\bf 146} (1992) 1.

\bibitem{JMN} M.~Jimbo, K.~Miki, T.~Miwa, A.~Nakayashiki, Phys. Lett. A {\bf 168} 
(1992) 256.

\bibitem{JMbook}
M.~Jimbo and T.~Miwa, {\it Algebraic Analysis of Solvable Lattice Models}, 
CBMS Regional Conference Series in Mathematics 
vol.{\bf 85}, American Mathematical Society, Providence, 1994.

\bibitem{KIEU}
V.~Korepin, A.~Izergin, F.~Essler and D.~Uglov, Phys. Lett. A {\bf 190} (1994) 182.

\bibitem{JM} M. Jimbo, T. Miwa, J.Phys. A {\bf 29} (1996) 2923.

\bibitem{ns} A. Nakayashiki, F.A. Smirnov, Commun. Math. Phys. {\bf 217} (2001) 623.

\bibitem{n} A. Nakayashiki, {\it ``On the Cohomology of Theta Divisor of 
Hyperelliptic Jacobian"},  Contemporary mathemathics, vol 309, in Integrable 
systems, topology and physics, M. Guest et al. ed., American Mathematical Society, 
Providence, 2002.

\bibitem{Maillet1}
N.~Kitanine, J.M.~Maillet, V.~Terras, Nucl. Phys. B {\bf 554} (1999), 647.

\bibitem{Maillet2}
N.~Kitanine, J.M.~Maillet, V.~Terras, Nucl. Phys. B {\bf 567} (2000), 554.

\bibitem{bk1} H.E. Boos, V.E. Korepin, J. Phys. A {\bf 34} (2001) 5311.

\bibitem{bk2} H.E.~Boos, V.E.~Korepin, 
{\it ``Evaluation of integrals representing correlators
in XXX Heisenberg spin chain"} in. MathPhys Odyssey 2001, 
Birkh\"auser, Basel, (2001) 65.

\bibitem{bks} H.E.~Boos, V.E.~Korepin, F.A.~Smirnov, 
Nucl. Phys.  {\bf B} 658 (2003) 417.

\bibitem{bks1} H.E.~Boos, V.E.~Korepin and F.A.~Smirnov, 
J. Phys. A: Math. Gen. {\bf 37} (2004) 323. 

\bibitem{bks2} H.E.~Boos, V.E.~Korepin and F.A.~Smirnov, 
{\it ``New formulae for solutions of quantum 
Knizhnik-Zamolodchikov equations on level $-4$
and correlation functions"},
hep-th/0305135, to appear in Moscow Mathematical Journal.

\bibitem{bks3} H.E.~Boos, V.E.~Korepin and F.A.~Smirnov,
{\it ``Connecting lattice and relativistic models via conformal
field theory"}, math-ph/0311020, Prog. Math. in press.

\bibitem{BKNS} H.E.~Boos, V.E.~Korepin, Y.~Nishiyama and M.~Shiroishi,
J. Phys. A: Math. Gen {\bf 35} (2002) 4443.

\bibitem{SSNT} K.~Sakai, M.~Shiroishi, Y.~Nishiyama, M.~Takahashi, 
Phys. Rev. E {\bf 67} (2003) 065101.

\bibitem{KSTS1} G.~Kato, M.~Shiroishi, M.~Takahashi, K.~Sakai, 
J. Phys. A: Math. Gen. {\bf 36} (2003) L337. 

\bibitem{TKS} M.~Takahashi, G.~Kato, M.~Shiroishi, 
J. Phys. Soc. Jpn. {\bf 73} (2004) 245. 

\bibitem{KSTS2} G.~Kato, M.~Shiroishi, M.~Takahashi, K.~Sakai, 
J. Phys. A: Math. Gen. {\bf 37} (2004) 5097. 

\bibitem{BJMST} H.~Boos, M.~Jimbo, T.~Miwa, F.~Smirnov and Y.~Takeyama,
{\it ``A recursion formula for the correlation functions of an inhomogeneous 
XXX model"}, hep-th/0405044. 

\bibitem{MT} N.~Muramoto and M.~Takahashi,
J. Phys. Soc. Jpn. {\bf 68} (1999) 2098.
\end{thebibliography}
\end{document}